\documentclass[12pt]{article}

\pdfoutput=1 
\usepackage{graphicx} 
\usepackage{color}                                                                                                                                                                                                                                                                                                                                                                                                                                                                                                                                                                                                                                                                                                                                                                                                                                                                                                                                                                                                                        
\usepackage{cancel,tabularx,moreverb,fancybox,amsmath,float,bm,braket,slashbox,txfonts,amssymb,bm,accents}
\usepackage[top=30truemm,bottom=30truemm,left=25truemm,right=25truemm]{geometry}
\usepackage{latexsym}
\usepackage{here}
\usepackage{cite}
\usepackage{hyperref}

\newcommand{\ex}[1]{\mathrm{e}^{#1}}

\newcommand{\pa}[1]{\left(#1 \right)}

\newcommand{\BR}[1]{\Biggl[#1 \Biggr]}

\newcommand{\kagi}[1]{\lbrack#1 \rbrack}
\newcommand{\bb}[1]{\mathbb{#1}}

\newcommand{\ca}[1]{\mathcal{#1}}

\newcommand{\abs}[1]{\left|#1\right|}
\newcommand{\ave}[1]{\langle #1\rangle}
\newcommand{\ar}[1]{\xrightarrow[#1]{}}
\newcommand{\pd}[1]{\frac{\partial}{\partial #1}}

\newcommand{\fr}{\frac}
\newcommand{\s}[1]{\sqrt{#1}}

\def\be{\begin{equation}}
\def\ee{\end{equation}}
\def\ba{\begin{eqnarray}}
\def\ea{\end{eqnarray}}

\def\gsim{{\gtrsim}}
\def\lsim{{\lesssim}}
\def\del{{\partial}}
\def\m{{\mu}}
\def\la{{\lambda}}
 \def\w{{\omega}}
 
 \def\ep{{\epsilon}}
 \def\d{{\delta}}

 \def\a{{\alpha}}
 
 \def\l{{\lambda}}
 \def\G{{\Gamma}}
 \def\D{{\Delta}}

 \def\b{{\beta}}
 \def\e{{\epsilon}}

 \def\p{\partial}

\def\tr{{\text{tr}}}

\def\bw{{\bar{w}}}
\def\dd{{\mathrm{d}}}
\def\sgn{{\text{sgn}}}

  \makeatletter
    
    \@addtoreset{equation}{section}
  \makeatother

\begin{document}

\begin{titlepage}
\thispagestyle{empty}

\begin{flushright}
YITP-18-106
\\

\end{flushright}

\bigskip

\begin{center}
\noindent{{\large \textbf{
Light Cone Bootstrap in General 2D CFTs\\
and \\
Entanglement from Light Cone Singularity
}}}\\
\vspace{2cm}
Yuya Kusuki
\vspace{1cm}

{\it
Center for Gravitational Physics, \\
Yukawa Institute for Theoretical Physics (YITP), Kyoto University, \\
Kitashirakawa Oiwakecho, Sakyo-ku, Kyoto 606-8502, Japan.
}
\vskip 2em
\end{center}

\begin{abstract}
The light cone OPE limit provides a significant amount of information regarding the conformal field theory (CFT), like the high-low temperature limit of the partition function.
We started with the light cone bootstrap in the {\it general} CFT ${}_2$ with $c>1$. For this purpose, we needed an explicit asymptotic form of the Virasoro conformal blocks in the limit $z \to 1$, which was unknown until now. In this study, we computed it in general by studying the pole structure of the {\it fusion matrix} (or the crossing kernel). Applying this result to the light cone bootstrap, we obtained the universal total twist (or equivalently, the universal binding energy) of two particles at a large angular momentum. In particular, we found that the total twist is saturated by the value $\frac{c-1}{12}$ if the total Liouville momentum exceeds beyond the {\it BTZ threshold}. This might be interpreted as a black hole formation in AdS${}_3$.

As another application of our light cone singularity, we studied the dynamics of entanglement after a global quench and found a Renyi phase transition as the replica number was varied. We also investigated the dynamics of the 2nd Renyi entropy after a local quench. 

We also provide a universal form of the Regge limit of the Virasoro conformal blocks from the analysis of the light cone singularity. This Regge limit is related to the general $n$-th Renyi entropy after a local quench and out of time ordered correlators.
\end{abstract}
 
\end{titlepage}

\restoregeometry

\tableofcontents
\section{Introduction \& Summary}

Two-dimensional conformal field theories (2D CFTs) have an infinite dimensional symmetry, the so-called Virasoro symmetry. This symmetry leads to the fact that 2D CFTs are only specified by a central charge, spectrum of primary operators, and operator product expansion (OPE) coefficients of the primary operators in the spectrum. Owing to this simplification, 2D CFTs offer ideal avenues for exploring quantum field theories. They are also investigated as the key to understanding quantum mechanics in AdS by using the AdS/CFT correspondence. However, despite several decades of studies, no criterion has been devised to classify CFTs and only a few models (for example, minimal models) are classified; in other words, we do not know how to identify which CFT data are consistent with the modular invariance.

One important tool to determine which CFT data are consistent is {\it conformal bootstrap}, or equivalently, the crossing symmetry 
\cite{Belavin1984,Ferrara1973,Polyakov1974,Rattazzi2008}.
The conformal bootstrap equation originates from the OPE associativity.
\begin{equation}
\fr{1}{x_{12}^{2\D_1}x_{34}^{2\D_2}}\sum_{p}P^{11,22}_{\tau_p,l_p} g^{11,22}_{\tau_p,l_p}(u,v)=\fr{1}{\pa{x_{14}x_{23}}^{\D_1+\D_2}}\pa{\fr{x_{24}}{x_{12}}}^{\D_{12}}\pa{\fr{x_{13}}{x_{12}}}^{\D_{12}}\sum_{p}P^{12,12}_{\tau_p,l_p} g^{12,12}_{\tau_p,l_p}(v,u),
\end{equation}
where $u,v$ are the cross ratios, and $g^{ij,kl}_{\tau,l}(u,v)$, $P^{ij,kl}_{\tau,l}$ are the conformal blocks and their coefficients.
This equation relates different OPE coefficients, and therefore, nontrivial requirements for CFT data can be obtained using the conformal bootstrap equation. In particular, the limit $u,1-v \to 0$ (or equivalently, $z, \bar{z} \to 0$) relates high-energy physics (i.e. information about the spectrum at large conformal dimensions) to the vacuum contribution. From this {\it high-low temperature duality}, 
\footnote{The conformal bootstrap equation of a particular four-point function in the limit $z,\bar{z} \to 0$ is related to the modular bootstrap in the high-low temperature limit $\beta \to 0$ through a map $\ex{-\beta}=\ex{-\pi\fr{K(1-z)}{K(z)}}$, where $K(z)$ is the elliptic integral of the first kind. For this reason, we will call the limit $z,\bar{z} \to 0$ as the {\it high-low temperature limit}.}
we can evaluate, for example, the density of states at large conformal dimensions, the so-called Cardy formula \cite{Cardy1986a}. Apart from this famous consequence, we also have several results from this duality \cite{Kraus2016, Cardy2017, Hikida2018, Romero-Bermudez2018, Brehm2018}.

Recently, another kinematic limit, {\it light cone limit} $u \ll v \ll 1$ (or equivalently $z\ll 1-\bar{z} \ll 1$), has attracted considerable interests in higher-dimensional CFTs \cite{Fitzpatrick2013a, Komargodski2013,Alday2015,Kaviraj2015,Kaviraj2015a, Alday2017a, Simmons-Duffin2017,Alday2017,Sleight2018}. Unlike the limit $z,\bar{z} \to 0$, the light cone limit relates the vacuum blocks to the OPE at large spin. Therefore, the light cone bootstrap reveals the structure of the OPE in the large spin limit. In fact, the light cone bootstrap imposes a condition such that in the CFT${}_{d\geq 3}$ with two scalar operators $\phi_A$ and $\phi_B$, there must exist operators with twist \cite{Fitzpatrick2013a, Komargodski2013}
\begin{equation}
\tau= \D_A+\D_B+2n+\gamma_{AB}(n,l) \ \ \ \ \ \text{for any } n\in \bb{Z}_{\geq0},
\end{equation}
and in particular, the anomalous dimension $\gamma_{AB}(n,l) \to 0$ as $l \to \infty$. In this way, the light cone limit provides substantial information about the spectrum.

In this paper, we give the following results:
\\
\\
\\
{\large [Light Cone Singularity]}
In 2D CFTs, the task of solving the light cone bootstrap is much more difficult than that in CFT${}_{\geq 3}$, as pointed out in \cite{Fitzpatrick2014}. The most difficult point is to evaluate the limit $z\to1$ of the Viraosoro conformal blocks (which we will call {\it light cone limit Virasoro blocks} in the following). At present, there is no exact asymptotic formula for the light cone limit of Virasoro blocks, let alone their explicit form. However, in this study, we accomplished to provide the asymptotic form of the light cone limit Virasoro blocks in any unitary CFT with $c>1$. The method for accomplishing this is to study the structure of poles in the fusion matrix $ {\bold F}_{\a_s, \a_t} $, which are invertible fusion transformations between $s$ and $t$-channel conformal blocks.
\begin{equation}
\begin{aligned}
\ca{F}^{21}_{34}(h_{\a_s}|z)=\int_{\bb{S}} \dd \a_t {\bold F}_{\a_s, \a_t} 
   \left[
    \begin{array}{cc}
    \a_2   & \a_1  \\
     \a_3  &   \a_4\\
    \end{array}
  \right]
  \ca{F}^{23}_{14}(h_{\a_t}|1-z),
\end{aligned}
\end{equation}
where $h_i=\a_i(Q-\a_i)$ and $c=1+6Q^2$.
Consequently, we can explicitly provide the light cone singularity of the conformal blocks. We then summarize the results, but before that, we will introduce some notations.
In this paper, we consider two types of Viraosoro conformal blocks. One is the Virasoro block for the correlator $\braket{O_B(\infty)O_B(1)O_A(z)O_A(0)}$ in the $O_A(z)O_A(0)$ OPE channel,
\newsavebox{\boxpa}
\sbox{\boxpa}{\includegraphics[width=160pt]{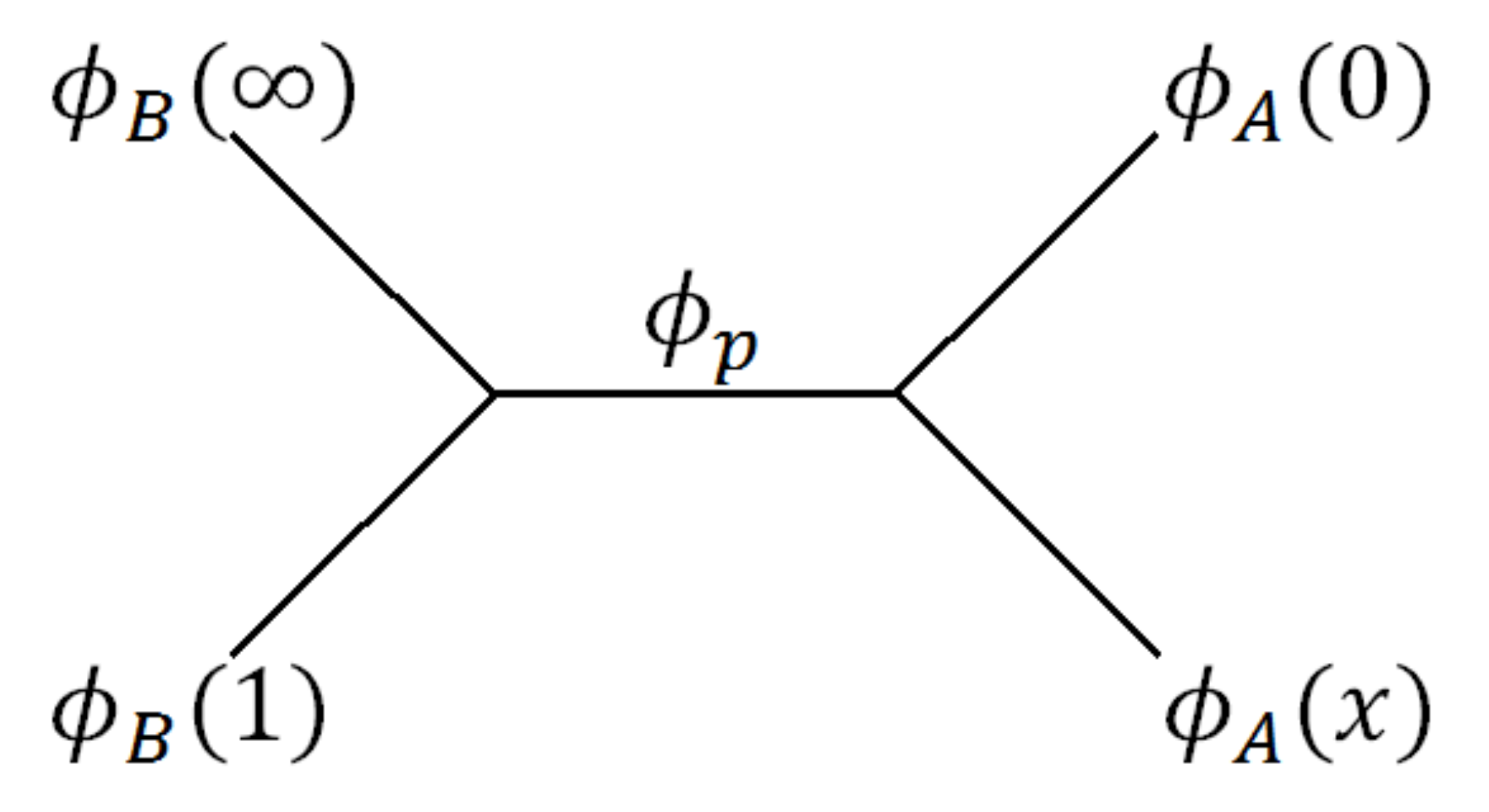}}
\newlength{\paw}
\settowidth{\paw}{\usebox{\boxpa}} 

\begin{equation*}
 \ca{F}^{AA}_{BB}(h_p|x) \equiv \parbox{\paw}{\usebox{\boxpa}},
\end{equation*}
which we call {\it AABB blocks}. The other type is {\it ABBA blocks}, which are defined as
\newsavebox{\boxpd}
\sbox{\boxpd}{\includegraphics[width=160pt]{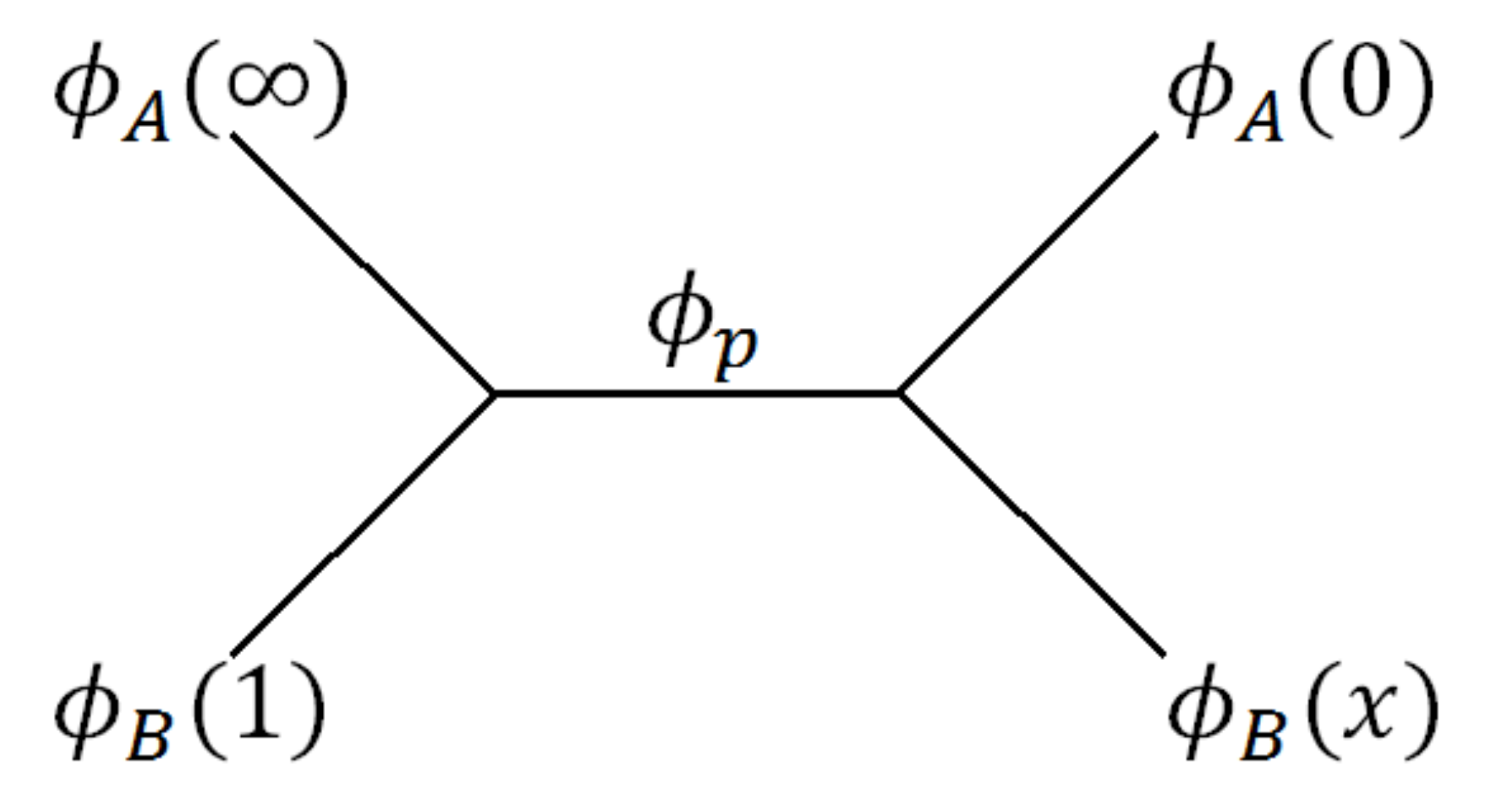}}
\newlength{\pdw}
\settowidth{\pdw}{\usebox{\boxpd}} 
\begin{equation*}
 \ca{F}^{BA}_{BA}(h_p|x) \equiv \parbox{\pdw}{\usebox{\boxpd}}.
\end{equation*}
The light cone singularities for the ABBA blocks is given by
\begin{equation}\label{eq:ABBAs}
\begin{aligned}
\ca{F}^{BA}_{BA}(h_{\a_s}|z)& \ar{z\to 1} \left\{
    \begin{array}{ll}
     (1-z)^{4h_A-2h_B-2Q\a_A}  ,& \text{if } \a_A<\fr{Q}{4}\ \text{and } \ \a_A<\a_B   ,\\
     (1-z)^{2h_B-2Q\a_B}  ,& \text{if } \a_B<\fr{Q}{4}\ \text{and } \ \a_B<\a_A   ,\\
     (1-z)^{\fr{c-1}{24}-2h_B}  ,& \text{otherwise } .\\
    \end{array}
  \right.\\
\end{aligned}
\end{equation}
The asymptotics of the AABB blocks are given by
\begin{equation}\label{eq:AABBs}
\begin{aligned}
\ca{F}^{AA}_{BB}(h_{\a_s}|z)& \ar{z\to 1} \left\{
    \begin{array}{ll}
     (1-z)^{-2\a_A\a_B}  ,& \text{if } \a_A+\a_B<\fr{Q}{2}\   ,\\
     (1-z)^{\fr{c-1}{24}-h_A-h_B}  ,& \text{otherwise }  .\\
    \end{array}
  \right.\\
\end{aligned}
\end{equation}
This is one of the main results presented in this paper. One interesting point is that we can find a transition of the blocks. However, in general, we cannot observe this transition in the behaviour of a four-point function because the approximation by only one (or few) block contribution appears if considering the limit $z,\bar{z} \to 0$ of the t-channel expansion. If one wants to find this transition in a correlator, then one has to find a quantity that can be evaluated by only one block contribution and in the limit $z \to 1$. In fact, it is given by the {\it light cone limit}
\begin{equation}
\braket{O_B(\infty) O_B(1) O_A(z,\bar{z}) O_A(0)} \ar{\text{light cone limit}} \ca{F}^{AA}_{BB}(0|z\to0)\overline{\ca{F}^{AA}_{BB}(0|\bar{z}\to1)}.
\end{equation}
Therefore, we can detect the transition in the light cone limit.
We want to emphasize that the transition point of the AABB blocks is characterised by the BTZ threshold $\a_{\text{BTZ}}=\fr{Q}{2}$, where $h_{\text{BTZ}}=\a_{\text{BTZ}}(Q-\a_{\text{BTZ}})=\fr{c}{24}$. Therefore, we expect that this {\it light cone transition} captures some important properties of holographic CFTs. In fact, one gravity interpretation of this transition can be obtained by the light cone bootstrap as we will explain later.

The detailed derivations of these singularities are provided in Appendix \ref{app:FM}.
In fact, this conclusion is supported by numerical computations \cite{Kusuki2018b,Kusuki2018} and proven in large $c$ CFTs \cite{Kusuki2018a}. We intend to emphasize that the fusion matrix approach does not rely on the assumption $c \to \infty$, unlike the Zamolodchikov monodromy method \cite{Kusuki2018a}; therefore, the conclusion (\ref{eq:AABBs}), (\ref{eq:ABBAs}) holds true not only for large $c$ CFTs but also for any unitary CFT with $c>1$. Note that our formula breaks down for the CFT data ($c<1$) of the minimal model, as explained in the final paragraph of Appendix \ref{subapp:FM}.

In addition, we discuss the sub-leading terms of the blocks in the light cone limit. In Appendix \ref{subapp:sub}, we investigate the other poles (sub-leading poles) in the fusion matrix and identify them as the sub-leading terms. In fact, the sub-leading singularities perfectly match the HHLL Virasoro blocks’ singularities in the heavy-light limit.
\\
\\
\\
{\large [Light Cone Bootstrap]}
In the main sections of this paper, we will apply this explicit form of light cone singularity to study the light cone bootstrap.
One of the main objectives is to understand the AdS${}_3$/CFT${}_2$ duality (for example, one concrete question is how Virasoro blocks can be derived from AdS gravity). Many recent developments pertaining to conformal blocks helps uncover information about holographic CFTs \cite{Fitzpatrick2013,Fitzpatrick2015,Fitzpatrick2017,Alkalaev2015,Hijano2015,Hijano2015a,Alkalaev2016,Asplund2015,Roberts2015,Alkalaev2016b, Chen2016, Alkalaev2016a, Hulik2017, Chen2017, Alkalaev2017, Maxfield2017,Alkalaev2018,Hikida2018c,Banerjee2018,Hikida2018a,Belavin2018, Hulik2018, Chen2018} . Similarly, it is expected that the light cone singularities revealed in this paper also lead to some interesting predictions in holographic CFTs.
In fact, we found that there must be a universal long-distance interaction between two objects at large spin for any unitary CFT with $c>1$ and without extra conserved current. 
Particularly, we found that the total twist of two particles at large spin is saturated by the BTZ threshold $\fr{c-1}{12}$ if the total Liouville momentum $\a_A+\a_B$ increases beyond the BTZ threshold $\a_{\text{BTZ}}=\fr{Q}{2}$.
\\
\\
\\
{\large [Entanglement]}
Apart from the light cone bootstrap, the light cone limit Viraosoro blocks also appear in many different situations.
In the rest of this paper, we discuss {\it entanglement} using the explicit asymptotic form of the light cone limit Virasoro blocks. In Section \ref{subsec:MILC}, we first consider the Renyi entanglement entropy for a special setup with a Lorentz boosted interval and an unboosted interval, as shown in Figure \ref{fig:AB}. In this case, the light cone limit naturally appears in the calculation. As a result, we found a Renyi phase transition as the replica number was varied. This implies that when we try to evaluate entanglement using the Renyi entropy, we must use the limit $n \to 1$ carefully. Note that the transition point $n_*$ is always above $n=1$; therefore, this Renyi phase transition does not contradict with the derivation of the holographic entanglement entropy formula \cite{Lewkowycz2013}. In Section \ref{subsec:DMI}, we consider the Renyi entanglement entropy for doubled CFTs, which was introduced by \cite{Hartman2013,Asplund2015a}. In this setup, the light cone limit non-trivially appears in the calculation, but is similar to that in Section \ref{subsec:MILC}. Therefore, the transition at $n_*$ can also be found in its Renyi entanglement entropy. We also predict that maximal scrambling is characterized by the lowest bound on the light cone singularity of a particular correlator; on the other hand, the quasi particle picture comes from the upper bound. It means that the light cone limit gives us information about scrambling. In section \ref{subsec:2ndREE}, we evaluate the dynamics of the 2nd Renyi entropy after a local quench. The result leads to the prediction that in a unitary CFT with $c>1$ and no extra currents, the heavier the operator used to create a local quench, the larger the 2nd Renyi entropy becomes; however, if its dimension exceeds the value $\fr{c-1}{32}$, the 2nd Renyi entropy is saturated. This might be related to instability and thermalisation. (See also  \cite{Caputa2014,He2014,Numasawa2016,Caputa2017,He2017,Guo2018}, which reveled the growth of the entanglement entropy after a local quench in other setups.) 
\\
\\
\\
{\large [Regge Limit Universality]}
Using the Zamolodchikov recursion formula, we can obtain the upper bound of the Regge limit singularity in the Virasoro blocks using the light cone singularity. In particular, above the BTZ threshold $\a_A+\a_B>\a_{\text{BTZ}}$, the asymptotic form of the Regge limit Virasoro blocks can be given by a universal formula. We will explain this in Section \ref{sec:Regge}. From this result, we can predict, for example, the late time behaviour of out of time ordered correlators (OTOCs)
\footnote{The time evolution of OTOCs at late time is also discussed in \cite{Caputa2016,Caputa2017a,Perlmutter2016,Gu2016,Fan2018}.}
 and general $n$-th Renyi entropy after a local quench.
\\
\\
\\
Besides these main contents, we give the light cone modular bootstrap in Appendix \ref{app:modular} and also we will rewrite our light cone bootstrap in this context. We give the details of the Zamolodchikov recursion relation in Appendix \ref{app:recursion}.

At the end of this section, we would like to emphasize the interesting points of our results.
\begin{description}
\item[Beyond test mass limit]
In the light cone limit, we can calculate Virasoro conformal blocks beyond the test mass limit $\fr{h_L}{c}\ll1$. It might give a key to understanding dynamics of {\it multiple} deficit angle in AdS${}_3$.

\item[BTZ thershold]
The light cone singularity undergoes a transition at the BTZ thershold. This suggests that this {\it light cone transition} captures information about holographic CFTs in some sense, like the Hawking-Page transition. (One interpretation is obtained by the light cone bootstrap.)

\item[Liouville CFT \& 2+1 Gravity]
Interestingly, our result from the bootstrap equation suggests that in a particular regime, 2+1 gravity is non-trivially related to Liouville CFT (see Figure \ref{fig:potential}), which cannot be observed in the test mass limit.
 We expect that this gives new insights into the relation between Liouville CFT and 2+1 gravity (see the end of Appendix \ref{subapp:relation}).
\end{description}

\section{Light Cone Bootstrap}

Let us consider a unitary CFT with two primary operators $O_A$ and $O_B$ in the following.
In general, the operators appearing in the OPE between $O_A$ and $O_B$ have large anomalous dimensions originating from its interactions.
However, the crossing symmetry implies that the large spin limit simplifies the structure of the OPE. In $d\geq 3$ unitary CFTs, the large $l$ primary operators in the OPE have a twist 
\begin{equation}\label{eq:twist3d}
\tau_n=\tau_A+\tau_B+2n  \ \ \ \ \ \text{ for any } n\in\bb{Z}_{\geq 0}.
\end{equation}
This is proved in \cite{Alday2007,Fitzpatrick2013a, Komargodski2013} using the light cone bootstrap.

The key point is the existence of a twist gap between the vacuum and minimal twist. The unitarity imposes the following bounds on a twist spectrum (except for vacuum):
\begin{equation}
\begin{aligned}
\tau \geq\left\{
    \begin{array}{ll}
      \fr{d-2}{2} ,& \text{if } l=0  ,\\
      d-2 ,& \text{otherwise }   .\\
    \end{array}
  \right.\\
\end{aligned}
\end{equation}
This leads to the separation of the identity conformal block from the other contributions in the bootstrap equation
as the global blocks $g_{\tau,l}^{ij,kl}(u,v)$ have the singularity $u^{\fr{\tau}{2}} v^{\fr{1}{2}\D_{ij}}$ with $\D_{ij}=\D_i-\D_j$ at $u,v \to 0$.  
In particular, in the light cone limit, the identity provides the dominant contributions to one hand side ($t$-channel expansion) in the bootstrap equation as follows:
\begin{equation}
u^{-\fr{1}{2}(\D_A+\D_B)} \underset{u\ll v\ll1}{\sim} v^{-\fr{1}{2}(\D_A+\D_B)} u^{-\fr{1}{2}\D_{12}}\sum_{\tau,l} P_{\tau,l}g_{\tau,l}(v,u), 
\end{equation}
where the sum is taken over the primary operators of twist $\tau=\D-l$ and spin $l$ in the OPE between $O_A$ and $O_B$, and parameters $u,v$ are the cross ratios defined by
\begin{equation}
u=\pa{\fr{x_{12} x_{34}}{x_{24}x_{13}}}^2, \ \ \ \ v=\pa{\fr{x_{14} x_{23}}{x_{24}x_{13}}}^2, 
\end{equation}
with $x_{ij}=x_i-x_j$. In this equation, it can be observed that, to match the $v$-dependence of both sides, there must exist the operators with twist (\ref{eq:twist3d}).

However, in 2D CFTs, the twist bound is given by $\tau\geq0$, which means that the identity contribution cannot be separated from the other contributions. Therefore, the above process does not work in 2D CFTs.
Even though there are many contributions to the zero-twist part, we can incorporate the contributions into the Virasoro conformal blocks; thus, using the Virasoro algebra, we can investigate the large spin structure even in 2D CFTs.
From this background, it is interesting to investigate what is predicted from the light cone bootstrap in 2D CFTs.
We will discuss it in this section.
\footnote{In fact, the light cone bootstrap in 2D CFTs has been studied in \cite{Fitzpatrick2014}; however, this study focused on only the semiclassical limit as the study relied on using HHLL Virasoro blocks. Now that we have the most general blocks in the light cone limit, we will consider more general unitary CFTs and external operators.}

Before moving on to the light cone bootstrap, we will interpret this statement in terms of AdS.
The operators at large $l$ correspond to states with large angular momentum in AdS, and thus two-particle states at large $l$ are orbiting a common centre with a large angular momentum. At this stage, it is naturally expected that at large $l$, these two particles are well separated, as the interactions between two objects become negligible at large angular momentum. It means that the anomalous dimension of the two-object state should vanish (see Figure \ref{fig:binding}).

\subsection{Lowest Twist Operator at Large $l$}

In 2D CFTs, the conformal bootstrap equation can be given in terms of Virasoro conformal blocks as
\begin{equation}\label{eq:bootstrap}
\sum_p C_{12p}C_{34p} \ca{F}^{21}_{34}(h_p|z)\overline{\ca{F}^{21}_{34}}(\bar{h}_p|\bar{z})=\sum_p C_{14p}C_{23p} \ca{F}^{23}_{14}(h_p|1-z)\overline{\ca{F}^{23}_{14}}(\bar{h}_p|1-\bar{z}),
\end{equation}
where $C_{ijk}$ are OPE coefficients and $\ca{F}^{ij}_{kl}(h_p|z)$ are conformal blocks, which are usually expressed using the Feynman diagram as follows:

\newsavebox{\boxpc}
\sbox{\boxpc}{\includegraphics[width=130pt]{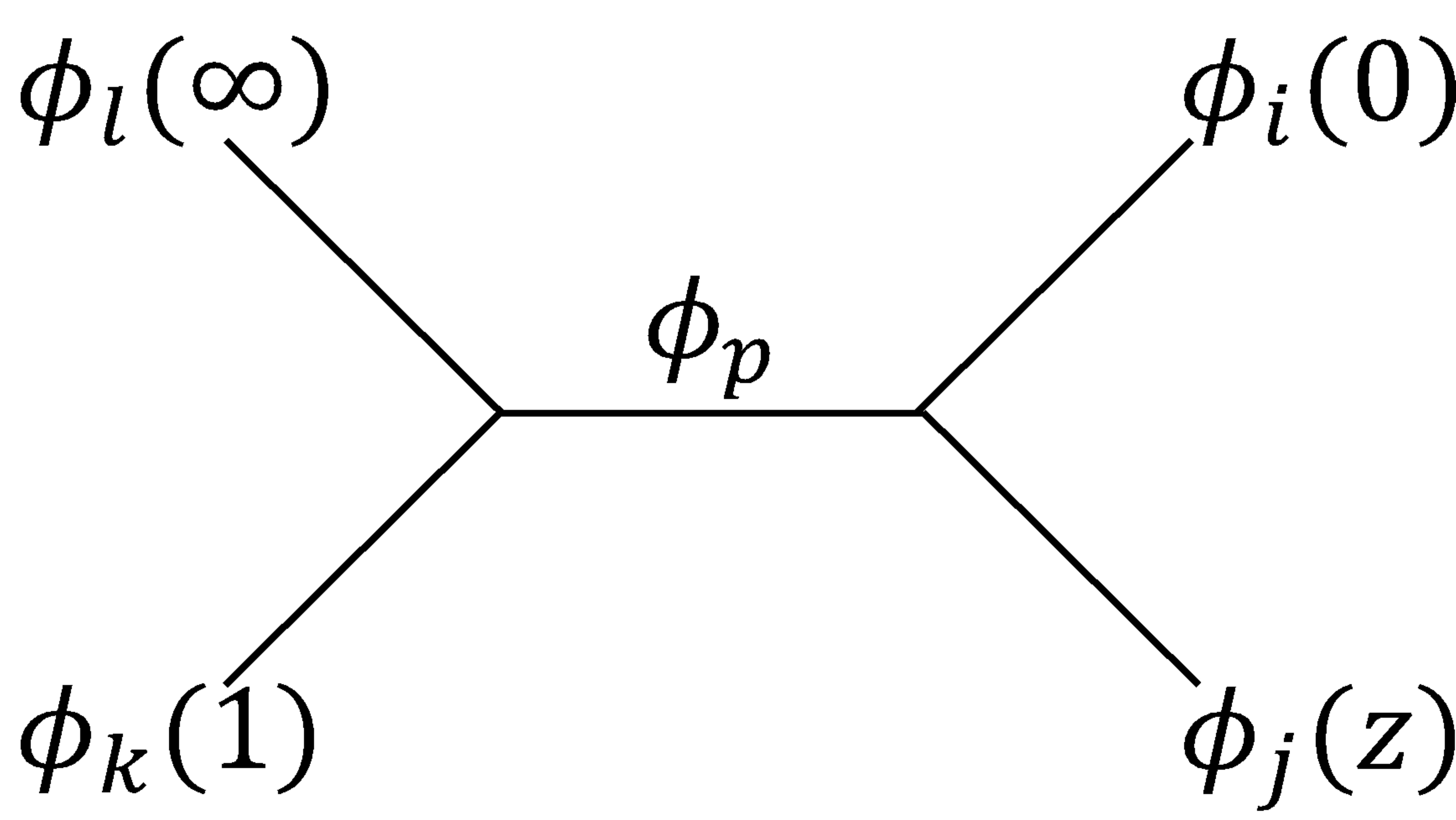}}
\newlength{\pcw}
\settowidth{\pcw}{\usebox{\boxpc}} 

\begin{equation*}
 \ca{F}^{ji}_{kl}(h_p|z) \equiv \parbox{\pcw}{\usebox{\boxpc}}.
\end{equation*}
We are interested in the light cone limit $z \ll 1-\bar{z} \ll 1$ of this equation. In the following, we assume that there are no additional continuous global symmetries apart from the Virasoro symmetry. Under this assumption, the light cone limit of the left hand side of (\ref{eq:bootstrap}) can be approximated using the vacuum Virasoro block. As the global block expansion is more useful to illustrate the asymptotics of the right hand side in the light cone limit, we re-express the right hand side using global blocks. As a result, the bootstrap equation in the light cone limit reduces to 
\begin{equation}
\ca{F}^{AA}_{BB}(0|z)\overline{F^{AA}_{BB} (0|\bar{z})}\simeq \pa{1-z}^{-h_A-h_B}  \pa{1-\bar{z}}^{-\bar{h}_A-\bar{h}_B}  \sum_{\tau,l}P_{\tau,l}g_{\tau,l}(z,\bar{z}),
\end{equation}
where $g_{\tau,l}(z,\bar{z})$ is the global block and $P_{\tau,l}$ is the conformal block coefficient. In this limit, we are interested in the large $l$ global blocks, which are given by a simple approximated form as
\begin{equation}
g_{\tau,l}(z,\bar{z}) \sim 2^{\tau+2l} (1-\bar{z})^{\fr{\tau}{2}} z^{\fr{1}{2}\D_{AB}}\s{\fr{l}{\pi}}K_{\D_{AB}}(2l\s{z}),
\end{equation}
where $\D=h+\bar{h}$, $\D_{ij}=\D_i-\D_j$, and $K_\D (x)$ are modified Bessel functions (see more details in \cite{Fitzpatrick2013a}).

To proceed further, we have to determine the behaviour of the Virasoro blocks in the limit $\bar{z} \to 1$. Although no exact closed form of the Virasoro blocks is available, if we restrict ourselves to the heavy-light limit, the HHLL Virasoro blocks make it possible to study the bootstrap equation discussed in \cite{Fitzpatrick2014}.
The HHLL Virasoro blocks in the limit $\bar{z} \to 1$ lead to
\begin{equation}
\begin{aligned}
\ca{F}^{LL}_{HH}(0|z)\overline{F^{LL}_{HH} (0|\bar{z})} 
&=(1-z)^{h_L(\d-1)}\pa{\fr{1-(1-z)^\d}{\d}}^{-2h_L} (1-\bar{z})^{\bar{h}_L(\bar{\d}-1)}\pa{\fr{1-(1-\bar{z})^{\bar{\d}}}{\bar{\d}}}^{-2\bar{h}_L} \\
&\ar{z \ll 1-\bar{z} \ll 1}
     z^{-2h_L}\pa{1-\bar{z}}^{\bar{h}_L(\bar{\d}-1) },
\end{aligned}
\end{equation}
where $\d=\s{1-\fr{24}{c}h_H}$ and $\bar{\d}=\s{1-\fr{24}{c}\bar{h}_H}$. For simplicity, we assume the external operators to be $h_i\geq\bar{h}_i$, and therefore $\tau_i=2 \bar{h_i}$.
When comparing the $z$ dependence of the left- and right-hand sides, one can find that there must be an infinite number of large $l$ contributions on the right-hand side to reproduce the singularity $z^{-2h_L}$ on the left-hand side. Moreover, to reproduce the singularity $\pa{1-\bar{z}}^{\bar{h}_L(\bar{\d}-1) }$ on the left-hand side, there must be a contribution from  an infinite number of operators having increasing spin with $\tau \to \d \tau_L+\tau_H$. This is discussed in \cite{Fitzpatrick2014}.

In the above, we restricted our investigation to the HHLL limit because it is extremely difficult to study the Virasoro blocks in general.
Nevertheless, the light cone limit interestingly simplifies the structure of the Virasoro blocks. We can find the simplification in the large $c$ limit achieved using the recursion relation \cite{Kusuki2018,Kusuki2018b} and monodromy method \cite{Kusuki2018a}.
In fact, we can also evaluate the light cone limit of {\it general} Virasoro conformal blocks using the fusion matrix.
Here, we will only present the results from the fusion matrix and present the detailed calculation in Appendix \ref{app:FM}, as the calculation is complicated.

In the following, we introduce notations usually found in Liouville CFTs.
\begin{equation}\label{eq:notations}
c=1+6Q^2, \ \ \ \ \ Q=b+\fr{1}{b}, \ \ \ \ \ h_i=\a_i(Q-\a_i).
\end{equation}
We denote $\a_i$ as {\it Liouville momenta}. We have to mention that although we use the notations in Liouville CFTs, we never use relations that only hold in Liouville CFTs. The Liouville parameters are introduced for convenience.
According to the result (\ref{eq:FMresult2}) in Appendix \ref{app:FM}, the light cone singularity is given by 

\begin{equation}\label{eq:LCsingularity}
\begin{aligned}
\ca{F}^{AA}_{BB}(0|z)\overline{F^{AA}_{BB} (0|\bar{z})} \ar{z \ll 1-\bar{z} \ll 1} &\left\{
    \begin{array}{ll}
     z^{-2h_A}\pa{1-\bar{z}}^{ -2\bar{\a}_A \bar{\a}_B } ,& \text{if} \ \ \ \bar{\a}_A+\bar{\a}_B<\fr{Q}{2} , \\
    z^{-2h_A}\pa{1-\bar{z}}^{\fr{c-1}{24} -\bar{h}_A-\bar{h}_B }  ,& \text{ otherwise},\\
    \end{array}
  \right.\\
\end{aligned}
\end{equation}
where $\a_i=Q\fr{1-\s{1-\fr{24}{c-1}h_i}}{2}$ and $\bar{\a}_i=Q\fr{1-\s{1-\fr{24}{c-1}\bar{h}_i}}{2}$. To reproduce this light cone singularity from the right-hand side of the bootstrap equation, we must always have the operator in the OPE with twist. 
\begin{equation}\label{eq:tlowest}
\begin{aligned}
\tau_{\text{lowest}}&=\left\{
    \begin{array}{ll}
    \tau_A+\tau_B -4 \bar{\a}_A \bar{\a}_B  ,& \text{if } \bar{\a}_A+\bar{\a}_B<\fr{Q}{2}  ,\\
     \fr{c-1}{12}  ,& \text{otherwise },\\
    \end{array}
  \right.\\
\end{aligned}
\end{equation}
where we can re-express the Liouville momenta $\bar{\a}_i$ using the twist $\tau_i$ as $\bar{\a}_i=Q\fr{1-\s{1-\fr{12}{c-1}\tau_i}}{2}$.
In particular, if we expand $\a_A$ in $\tau_{\text{lowest}}$ at small $\fr{h_A}{c}$, the result exactly matches $\tau_{\text{lowest}} \to \d \tau_A+\tau_B$.
This twist $\tau_{\text{lowest}}$ gives the lower bound for the twist in the OPE at large $l$ because if there exists an operator with twist lower than $\tau_{\text{lowest}}$, the singularity arising from that twist is never reproduced by the left-hand side of the bootstrap equation.

\begin{figure}[t]
 \begin{center}
  \includegraphics[width=7.0cm,clip]{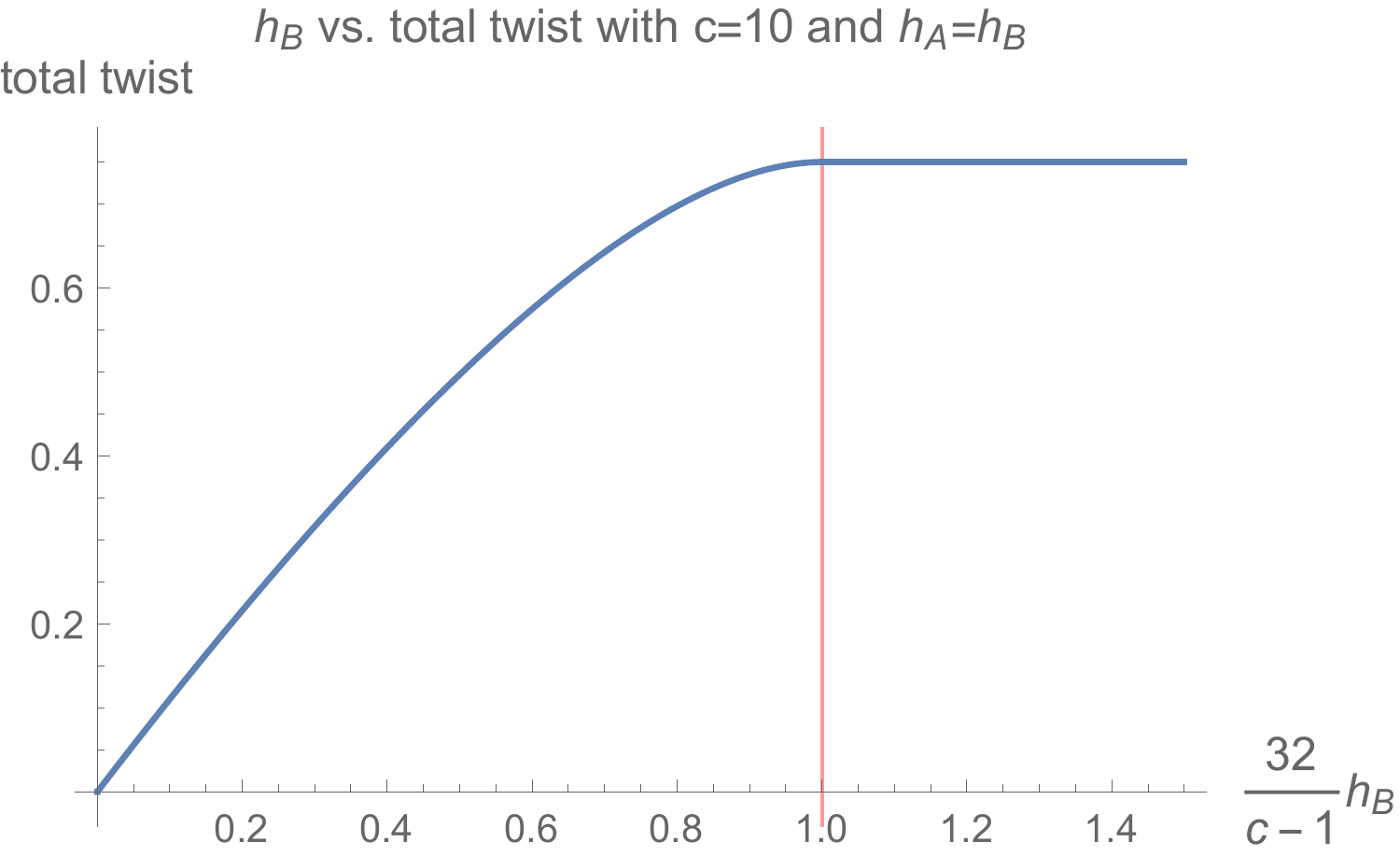}
  \includegraphics[width=7.0cm,clip]{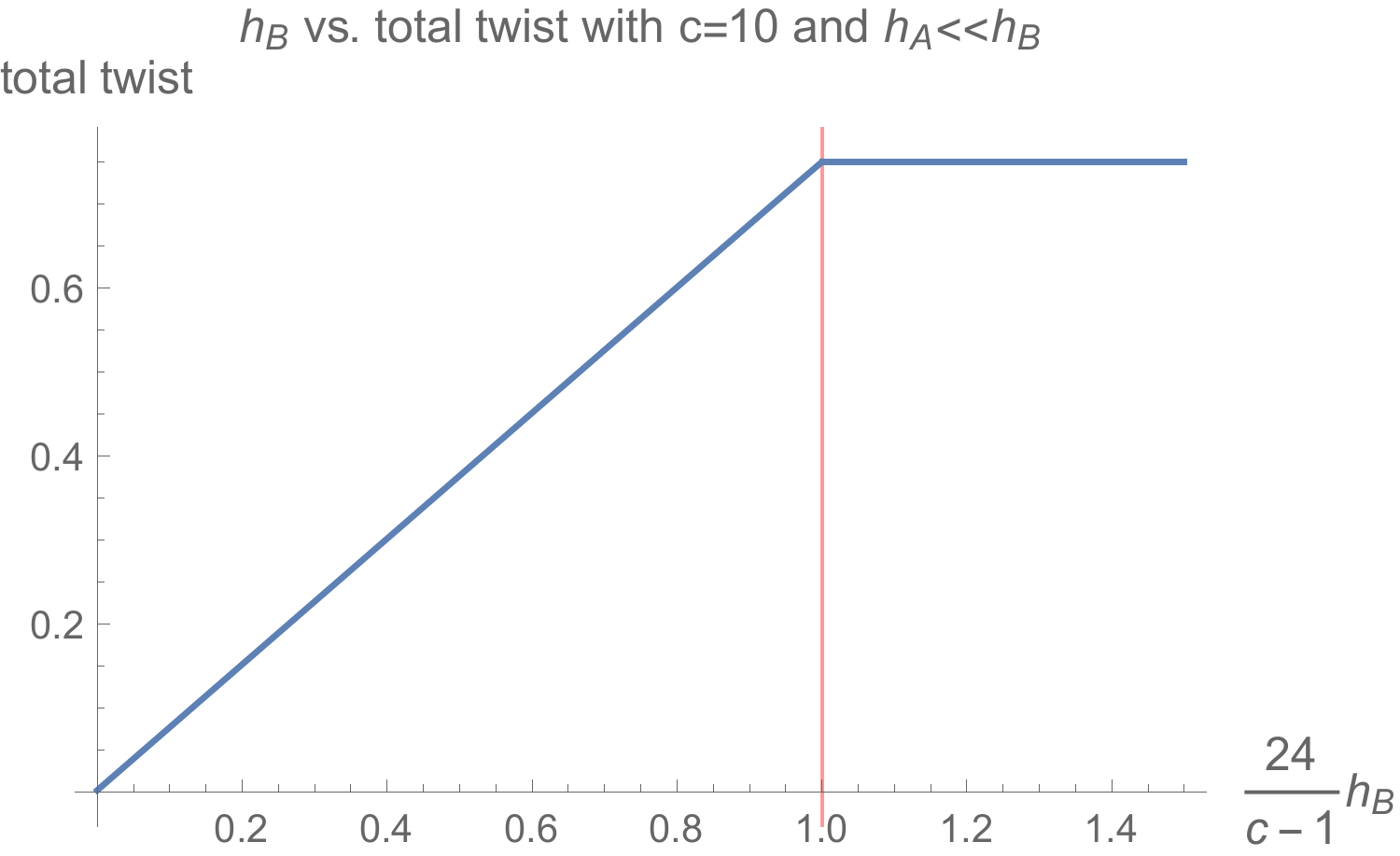}
 \end{center}
 \caption{The left figure shows the $h_A=h_B$ dependence of the total twist in the OPE between $O_A$ and $O_B$, and the right figure shows the $h_B$ dependence with $h_A \ll h_B$. From both figures, we can find saturation above $\a_A+\a_B=\fr{Q}{2}$.
In particular, for $h_A\ll h_B$, the transition (saturation) occurs at the BTZ mass threshold $\fr{c-1}{24}$. A similar phenomenon can be seen in the HHLL Virasoro blocks, known as {\it thermalisation}.}
 \label{fig:total}
\end{figure}

An interesting point is that if the total Liouville momentum $\tau_A+\tau_B$ increases beyond the BTZ momentum threshold $\a_{\text{BTZ}}=\fr{Q}{2}$
\footnote{The Liouville momentum $\a_{\text{BTZ}}=\fr{Q}{2}$ leads to the conformal dimension $h=\fr{Q^2}{4}=\fr{c-1}{24}$. This is just the BTZ mass threshold.},
the lowest twist in the OPE is given by a universal value $\fr{c-1}{12}$. Moreover, this universal twist equals the BTZ mass threshold. In other words, the total twist gradually increases unless the total Liouville momentum exceeds the BTZ mass threshold, and the total twist is saturated by the value of that threshold if the total Liouville momentum increases beyond it. This behaviour of the total twist is shown in Figure \ref{fig:total}. Here, one might face a contradiction to the thermalisation of the HHLL Virasoro blocks.
It is well known that the HHLL vacuum block can be interpreted as a two-point function on a thermal background if the mass of the heavy particle exceeds the BTZ threshold $h_H>\fr{c}{24}$. To explore this, let us consider the expression (\ref{eq:LCsingularity}). Considering the test mass limit $\fr{h_A}{c} \to 0$, the threshold $\bar{\a}_A+\bar{\a}_B<\fr{Q}{2}$ can be approximated using an inequality $h_B<\fr{c}{24}$. It means that the saturation of the singularity occurs exactly at the BTZ threshold
 $h_B=\fr{c}{24}$, as expected from the analysis of the HHLL Virasoro blocks. However, the singularity of the thermal correlator is considerably different from (\ref{eq:LCsingularity}). That is, above the BTZ threshold, the HHLL vacuum Virasoro block leads to the singularity
\begin{equation}
\begin{aligned}
\ca{F}^{LL}_{HH}(0|z)
&=(1-z)^{h_L(\d-1)}\pa{\fr{1-(1-z)^\d}{\d}}^{-2h_L}  \\
&\ar{z \to 1}
     \pa{1-z}^{-h_L },
\end{aligned}
\end{equation}
which is obviously different from the singularity in (\ref{eq:LCsingularity}).  This is perhaps due to the test mass limit $\fr{h_A}{c} \to 0$; here, the light cone limit could not commute with the test mass limit $h_L/c\to0$. It means that the back reaction of a probe actually results in a non-negligible universal interaction with a heavy particle in AdS${}_3$. Note that one can find the agreement of the transition points between HHLL blocks and light cone limit singularity, which might imply that the transition of the light cone singularity is also related to thermalisation, instability, or black hole formation. In fact, the assumptions $c \to \infty$ and no extra conserved current cause a CFT to be irrational, which is expected to be chaotic. Therefore, the assumptions might be appropriate for thermalisation to occur.
It would be interesting to explore this issue further.

We want to emphasize that our result (\ref{eq:tlowest}) holds not only for large $c$ CFTs but also for any unitary CFT with $c>1$ and no extra conserved currents. Note that the absence of the first condition $c>1$ destroys the light cone OPE structure (see the final paragraph of Appendix \ref{subapp:FM}), and the second condition is used to approximate the light cone limit of a correlator using the vacuum Virasoro block. Therefore, rational CFTs are an exception in our statement.

Finally, we would like to mention that the fact that the lowest twist is saturated by $\fr{c-1}{12}$ is consistent with the result from the light cone modular bootstrap. By using the modular symmetry, it is easy to show that there must be an infinite number of large spin primaries with twist accumulating to $\fr{c-1}{12}$ in any unitary CFT with $c>1$ and without extra currents (see Appendix \ref{app:modular} or \cite{Collier2016}). However, our statement (\ref{eq:tlowest}) is a little different because our result implies that there must be an infinite number of operators with twist $\fr{c-1}{12}$ at $ l \to \infty$ not only in the CFT, but also in the {\it OPE}. In this sense, our conclusion is more interesting than the result from the light cone modular bootstrap on a torus.
Possibly, we could show that the operator with twist $\fr{c-1}{12}$ predicted from the modular bootstrap actually comes from the OPE between  two operators with heavy total Liouville momentum ($\bar{\a}_A+\bar{\a}_B>\fr{Q}{2}$).

\subsection{Large $l$ Spectrum of Twist}\label{subsec:spectrum}

In the previous section, we derived the lowest twist at large $l$, but we would also like to determine the twist spectrum.
For this purpose, we need not only the leading conformal blocks but also the sub-leading contributions to the blocks.
 We omit the details of the calculation and only present the results; interested readers can refer to Appendix \ref{app:FM}, particularly \ref{subapp:sub} .

As the simplest example, we first consider the heavy-light limit. In the limit $c \to \infty$ with $\fr{h_H}{c} ,h_L$ fixed, the light cone asymptotics of the conformal blocks is given as (\ref{eq:subHHLL})
\begin{equation}
\ca{F}^{LL}_{HH}(h_p|z)\ar{z\to1}  \sum_{n \in \bb{Z}_{\geq0} } \ca{P}_n (1-z)^{\d(h_L+n)-h_L},
\end{equation}
where $\ca{P}_n$ are some constants. Therefore, the left-hand side of the bootstrap equation is 
\begin{equation}
\ca{F}^{LL}_{HH}(0|z)\overline{F^{LL}_{HH} (0|\bar{z})} 
\ar{z \ll 1-\bar{z} \ll 1}
     z^{-2h_L}\pa{\sum_{n \in \bb{Z}_{\geq0} } \ca{P}_n (1-\bar{z})^{\d(\bar{h}_L+n)-\bar{h}_L}}.
\end{equation}
To reproduce the $\bar{z}$ dependence of each term on the left-hand side, there must be at least one primary operator with twist.
\begin{equation}
\tau_n=\d(\tau_L+2n)+\tau_H \ \ \ \ \ \text{for any } n\in\bb{Z}_{\geq 0}.
\end{equation}
Let us consider the case when the condition $\fr{h_L}{c} \ll 1$ is relaxed.
In this case, the light cone limit of a four-point function is given as follows. If $\bar{\a}_A \bar{\a}_B<\fr{Q}{2}$,
\begin{equation}
\ca{F}^{AA}_{BB}(0|z)\overline{F^{AA}_{BB} (0|\bar{z})} 
\ar{z \ll 1-\bar{z} \ll 1}
     z^{-2h_A}\pa{\sum_{n \in \bb{Z}_{\geq0} } \ca{P}_n (1-\bar{z})^{-2\bar{\a}_A \bar{\a}_B +n\pa{1-\fr{2}{Q}(\bar{\a}_A+\bar{\a}_B)}}};
\end{equation}
otherwise,
\begin{equation}\label{eq:c/24}
\ca{F}^{AA}_{BB}(0|z)\overline{F^{AA}_{BB} (0|\bar{z})} 
\ar{z \ll 1-\bar{z} \ll 1}
     z^{-2h_A}(1-\bar{z})^{\fr{c-1}{24}-\bar{h}_A-\bar{h}_B}.
\end{equation}
As a result, the light cone bootstrap imposes a condition that there must exist operators with twist.
\footnote{
After our work appeared, similar work was done in \cite{Collier2018}. In \cite{Collier2018}, this spectrum is called {\it Virasoro Mean Fiald Theory}. (More higher corrections are given by (\ref{eq:general twist}).)
Actually, this large spin twist spectrum can be described by the fusion rule of the {\it Liouville CFT}, as explained in  Appendix \ref{subapp:relation}.
}

\begin{equation}\label{eq:generaltn}
\begin{aligned}
\tau_n&=\left\{
    \begin{array}{ll}
    \tau_A+\tau_B -4\bar{\a}_A \bar{\a}_B +2n\pa{1-\fr{2}{Q}(\bar{\a}_A+\bar{\a}_B)}  \ \ \ (\text{any } n \in \bb{Z}_{\geq0}) ,& \text{if } \bar{\a}_A+\bar{\a}_B<\fr{Q}{2}  ,\\
     \fr{c-1}{12}  ,& \text{otherwise }   .\\
    \end{array}
  \right.\\
\end{aligned}
\end{equation}
Note that an infinite tower comes from an infinite number of particular poles in the fusion matrix (see Appendix \ref{subapp:sub}).  This infiniteness relies on the assumption $c \to \infty$; therefore, when we exceed the semiclassical limit, there is only a finite tower of operators in the OPE, unlike (\ref{eq:generaltn}) in large $c$ CFTs.

In fact, for general unitary CFTs with $c>1$, the light cone limit of a four- point function can be approximated as follows:
\begin{equation}
\ca{F}^{AA}_{BB}(0|z)\overline{F^{AA}_{BB} (0|\bar{z})} 
\ar{z \ll 1-\bar{z} \ll 1}
     z^{-2h_A} \pa{\sum_{\substack{m,n \in \bb{Z}_{\geq0}\\ \text{where} \\ \bar{\a}_A+\bar{\a}_B+Q_{m,n}<\fr{Q}{2} }}
 \ca{P}_{m,n} (1-\bar{z})^{-2\bar{\a}_A \bar{\a}_B+\m_{m,n}}}, \ \ \ \ \  \text{if } \bar{\a}_A+\bar{\a}_B<\fr{Q}{2},
\end{equation}
where the correction $\m_{m,n}$ to the highest singular power law is defined using the Liouville notation (\ref{eq:notations}) and the notations  $\w_{m,n} \equiv \fr{2}{Q}(\bar{\a}_A+\bar{\a}_B+Q_{m,n})<1$, $Q_{m,n}=mb+\fr{n}{b}$ as
\begin{equation}
\m_{m,n}=\pa{mb+\fr{n}{b}} \pa{(1-\w_{m,n}+m)b+\fr{1-\w_{m,n}+n}{b}}.
\end{equation}
On the other hand, if $\bar{\a}_A+\bar{\a}_B>\fr{Q}{2}$, the light cone asymptotics for a correlator in general unitary CFTs are the same as those in (\ref{eq:c/24}).
The details of the calculation are given before equation (\ref{eq:fullAABB}).
We can thus conclude that in any 2D CFT with $c>1$ and no extra conserved currents, there must be operators with twist
\begin{equation}\label{eq:general twist}
\begin{aligned}
\tau_{m,n}&=\left\{
    \begin{array}{ll}
    \tau_A+\tau_B -4\bar{\a}_A \bar{\a}_B +2\m_{m,n}  \ \ \ (\text{any } m,n \in \bb{Z}_{\geq0} \text{ s.t. } \bar{\a}_A+\bar{\a}_B+Q_{m,n}<\fr{Q}{2} ) ,& \text{if } \bar{\a}_A+\bar{\a}_B<\fr{Q}{2}  ,\\
     \fr{c-1}{12}  ,& \text{otherwise }   .\\
    \end{array}
  \right.\\
\end{aligned}
\end{equation}
One can find that $\m_{m,n}$ differ by non-integer from one another, which means that each twist belongs to a different conformal family.
We would like to emphasize that this large spin twist spectrum can be described by the fusion rule of {\it Liouville CFT}, which is explained in Appendix \ref{subapp:relation}.

\begin{figure}[t]
 \begin{center}
  \includegraphics[width=10.0cm,clip]{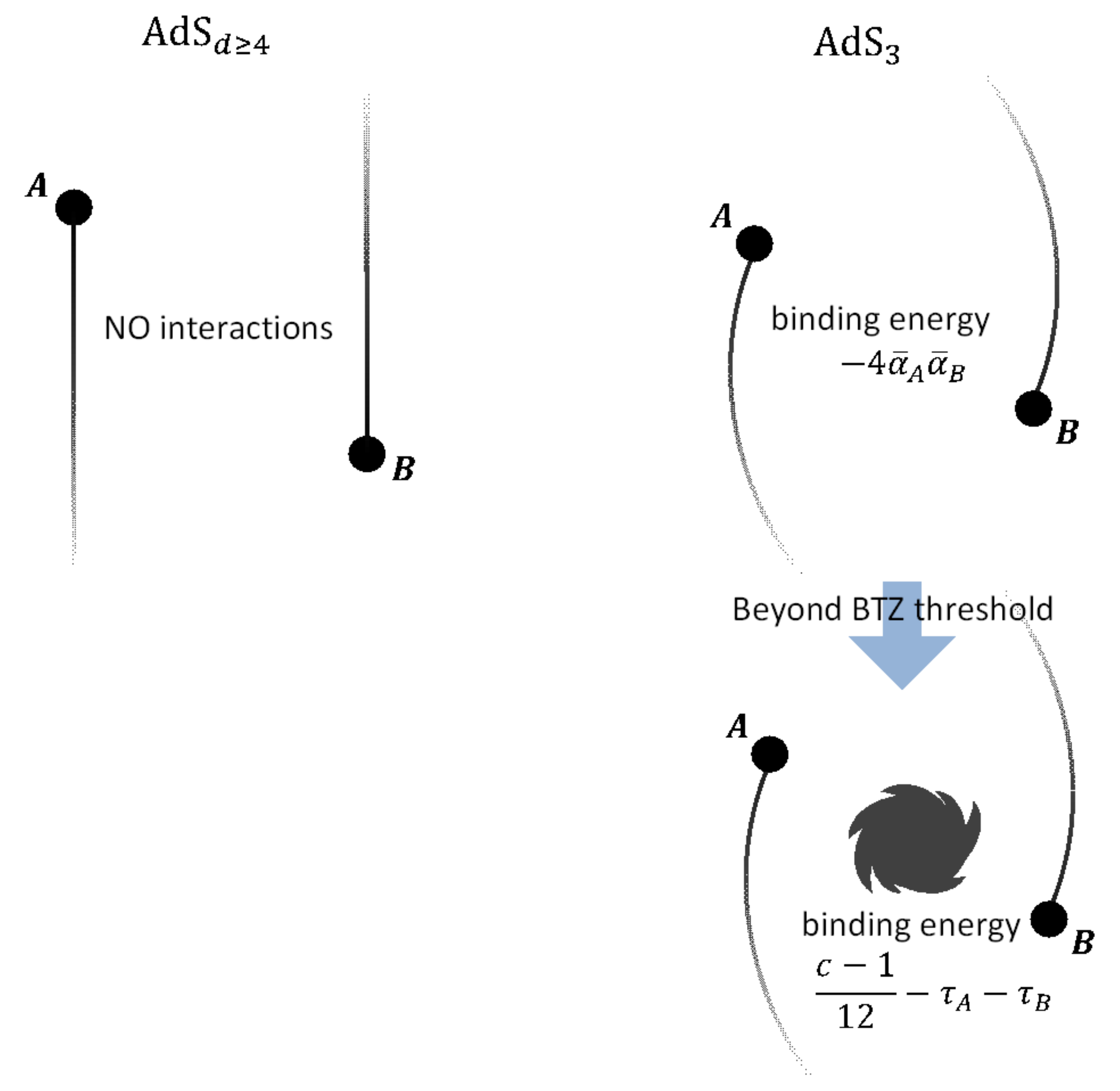}
 \end{center}
 \caption{This figure shows the implications of the light cone bootstrap on the nature of AdS. In AdS${}_{d\geq 4}$, the interactions between two objects become negligible at large angular momentum. On the other hand, in AdS${}_3$, there exists a universal binding energy $-4\bar{\a}_A\bar{\a}_B$ even at large angular momentum. If the total Liouville momentum is above the BTZ threshold $\a_{\text{BTZ}}$, the binding energy is given by $\fr{c-1}{12}-\tau_A-\tau_B$.
}
 \label{fig:binding}
\end{figure}

\begin{figure}[t]
 \begin{center}
  \includegraphics[width=12.0cm,clip]{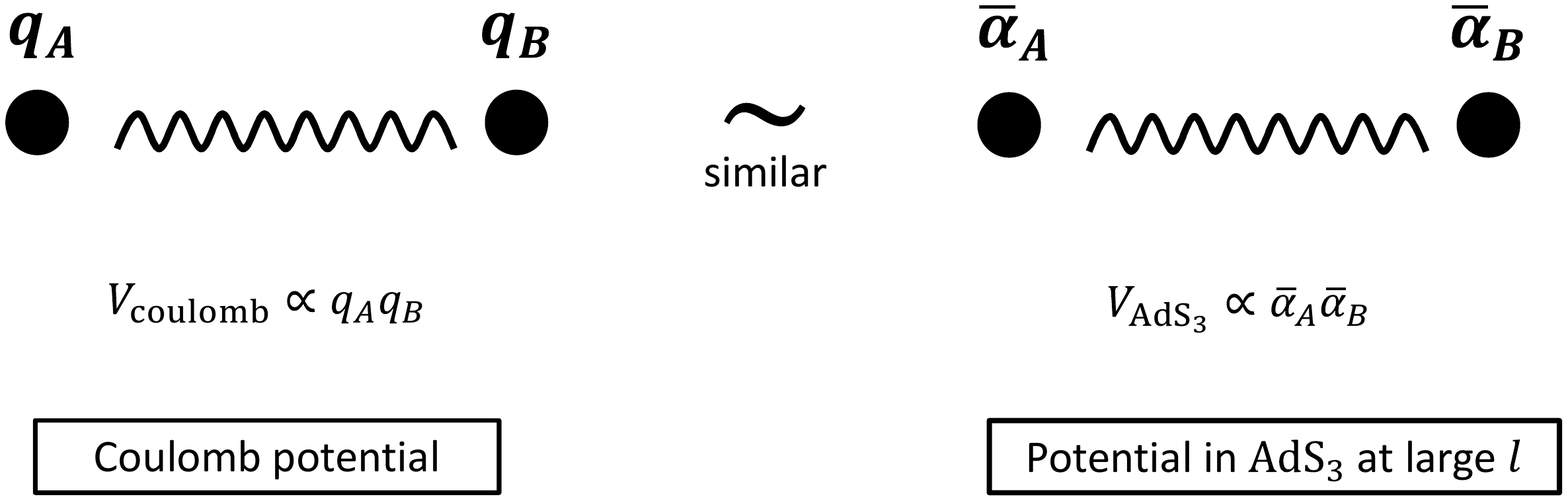}
 \end{center}
 \caption{Potential form in AdS$_3$ between two particles is similar to the form of Coulomb potential. Moreover, this form with Liouville momenta implies that the dynamics of the multiple deficit angle at large angular momentum is completely captured by Liouville CFT.}
 \label{fig:potential}
\end{figure}

This twist spectrum is considerably different from the prediction using the AdS interpretation, $\tau_n=\tau_A+\tau_B+2n$.
It means that the interactions between $O_A$ and $O_B$ never vanish even at large $l$. In other words, there is a universal anomalous dimension only in 2D CFTs. The reason is that gravitational interactions in AdS${}_3$ create a deficit angle, and their effect can be detected even at infinite separation.
When focusing on the test mass limit $\fr{h_A}{c} \to 0$, this long-distance effect can be interpreted clearly in AdS \cite{Fitzpatrick2014}. That is, the existence of a deficit angle in AdS${}_3$ leads to an energy shift 
\begin{equation}
\D_A \to \D_A \s{1-8G_N M} =\a \D_A,
\end{equation}
where we used the dictionary, $\D_B=M$ and $c=\fr{3}{2G_N}$. This energy shift is universal and never vanishes even at large angular momentum; therefore, it is natural that the corresponding twist in CFT${}_2$ is also shifted by
\begin{equation}
\tau_{\text{total}}=\tau_A+\tau_B \to \a\tau_A+\tau_B. 
\end{equation}
For general unitary CFTs with $c>1$, we can conclude that there is a universal binding energy between two objects at large $l$.
\begin{equation}\label{eq:Ebound}
\begin{aligned}
E_{\text{binding}}&\ar{l \to \infty}\left\{
    \begin{array}{ll}
     -4\bar{\a}_A \bar{\a}_B  ,& \text{if } \bar{\a}_A+\bar{\a}_B<\fr{Q}{2}  ,\\
     \fr{c-1}{12}-\tau_A-\tau_B  ,& \text{otherwise }   .\\
    \end{array}
  \right.\\
\end{aligned}
\end{equation}
This universal binding energy only exists in AdS${}_3$ and vanishes in AdS${}_{d\geq4}$, as shown in Figure \ref{fig:binding}.
This form of the binding energy below the BTZ threshold could be natural to some extent because similar to the gravity theory or electromagnetic theory, this form is characterized by the product of two Liouville momenta, where the Liouville momentum behaves like a {\it charge} (see Figure \ref{fig:potential}).
It would be interesting to see how this binding energy is obtained from the calculation in AdS${}_3$, determine why the Liouville momentum essentially appears in the expression of the binding energy and understand the physical meaning of this binding energy in general unitary CFTs with $c>1$. Further, we intend to understand what leads to saturation of the binding energy. We expect that this saturation is related to thermalisation and black hole formation.

\section{Entanglement and Light Cone Limit}

The light cone limit is useful in understanding entanglement, which is discussed in this section.

One useful measure of entanglement is entanglement entropy, which is defined as
\begin{equation}
S_A=-\tr \rho_A \log \rho_A,
\end{equation}
where $\rho_A$ is a reduced density matrix for a subsystem $A$, obtained by tracing out its complement. We will also discuss its generalization, called the Renyi entropy, which is defined as
\begin{equation}
S_A^{(n)}=\fr{1}{1-n} \log \tr \rho_A^n,
\end{equation}
and the limit $n\to1$ of the Renyi entropy defines the entanglement entropy $S_A$.
Here, to characterize entanglement precisely, we measure the Renyi entropy for two disconnected intervals $A \cup B$.
In particular, we consider the Renyi entropy in the light cone limit in Section \ref{subsec:MILC}. This has not been extensively studied as the explicit form of a four-point function in the light cone limit was unknown until our earlier studies \cite{Kusuki2018a,Kusuki2018,Kusuki2018b}.
\footnote{This setup is holographically studied in \cite{Kusuki2017}.}

The entanglement entropy for two disconnected intervals, or equivalently, the mutual information, is also useful to probe how entanglement spreads. 
The entanglement entropy for two disconnected intervals $S_{A\cup B}$ does not measure the entanglement of $A$ with $B$ but measures the entanglement of $A \cup B$ with its complement. Nevertheless, we can determine the entanglement of $A$ with $B$ from $S_{A\cup B}$ because a strong entanglement between $A$ and $B$ means that $A \cup B$ cannot be entangled with the rest; thus, one can find that $S_{A\cup B}$ is small if $A$ is highly entangled with $B$.

For example, Calabrese and Cardy studied the entanglement entropy for disconnected intervals to conclude that, after a global quench, entanglement spreads as if correlations were carried by free quasiparticles \cite{Calabrese2005a,Calabrese2009a}; this finding was refined in \cite{Asplund2015a} as the quasiparticle picture is only valid under some assumptions.

Interestingly, the light cone limit also appears in the study on the dynamics of entanglement as in \cite{Asplund2015a}. 
Based on this fact, we will discuss the dynamics of entanglement after a global quench in Section \ref{subsec:DMI}, \ref{subsec:DMI2} and after a local quench in Section \ref{subsec:2ndREE}.

\subsection{Mutual Information in Light Cone Limit} \label{subsec:MILC}

\begin{figure}[t]
  \centering
  \includegraphics[width=8cm]{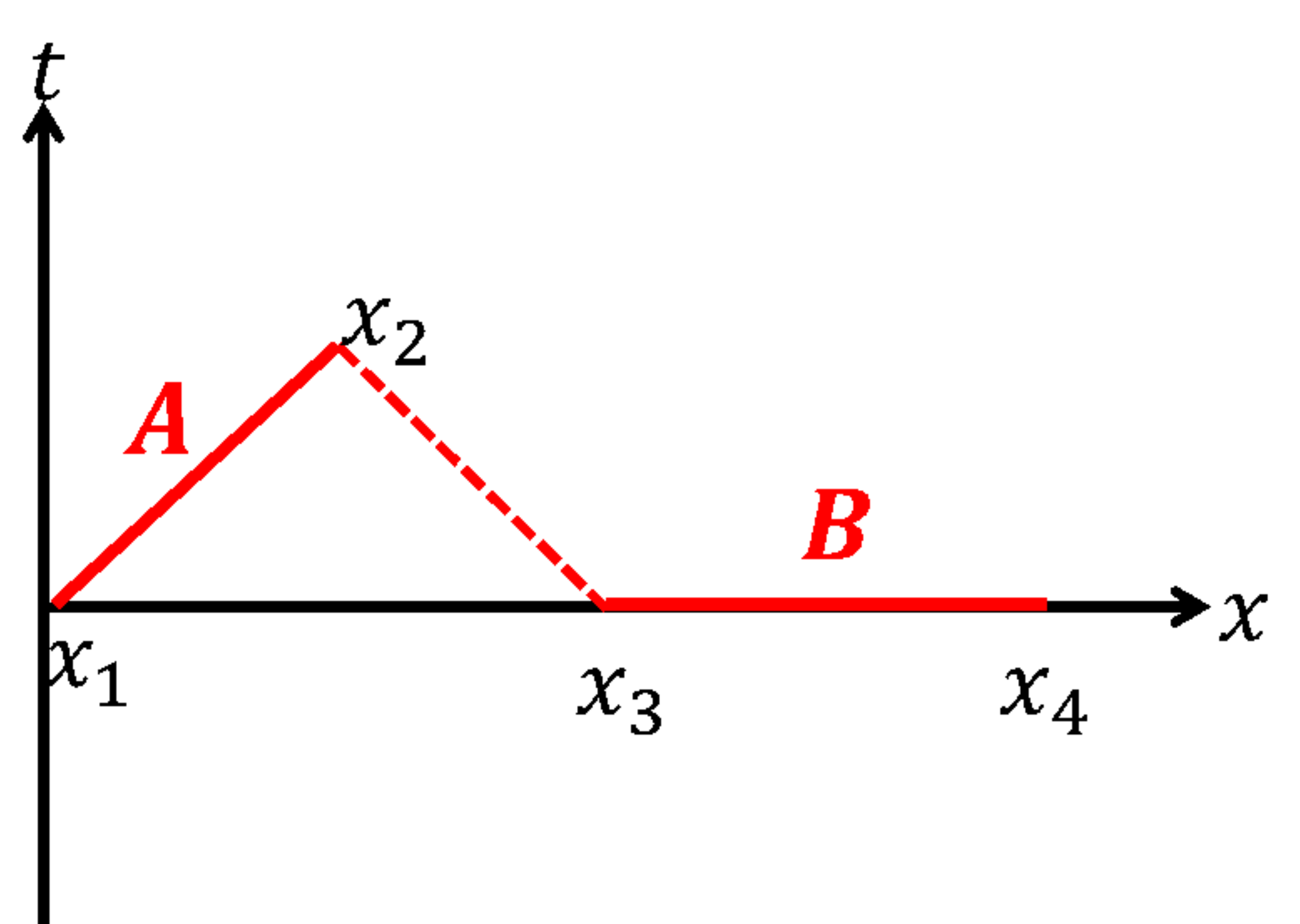}
  \caption{Setup of mutual information $I(A,B)$ between two intervals $A$ and $B$ when $A$ is infinitely boosted.}
\label{fig:AB}
  \end{figure}

In this section, we consider the light cone limit of the Renyi entanglement entropy $S^{(n)}_{A\cup B}$ for two intervals $A$ and $B$,
or the Renyi mutual information, which is given by 
\be\label{eq:MI}
 I^{(n)}(A : B)=S^{(n)}_{A}+S^{(n)}_{B}-S^{(n)}_{A\cup B} .
\ee
Let us choose $A$ and $B$ to be $[x_1,x_2]$ and $[x_3,x_4]$ in the 2D Lorentzian spacetime ${\bf R}^{1,1}$.
Note that the Lorentzian time $t$ and space $x$ are related to the complex coordinate as $z=x+it_E=x-t$.
In terms of the complex coordinate, the intervals are specified by the twist operators at
\begin{align}
z_1 &= \bar{z}_1 =  x_1, \ \  z_3 = \bar{z}_3 = x_3,  \ \ z_4 = \bar{z}_4 = x_4, \notag \\
z_2 &= x_2 -(-t) , \ \ \bar{z}_2 = x_2 + (-t),
\end{align}
where $x_1 < x_2 < x_3 < x_4$ and $t> 0$. Here, we consider a simple case where only $x_2$ goes away from the $t = 0$ slice.
The cross ratios are given by
\begin{align}
z
&= \frac{z_{12} z_{34}}{z_{13} z_{24}}
= \frac{(x_{21} - t ) x_{43}}{ x_{31} (x_{42} + t )}
=\frac{x_{21}^{-} x_{43}^{-}}{x_{31}^{-} x_{42}^{-} } , \notag \\
\bar{z}
&= \frac{\bar{z}_{12} \bar{z}_{34}}{\bar{z}_{13} \bar{z}_{24}}
= \frac{(x_{21} + t ) x_{43}}{ x_{31} (x_{42} - t)}
=1-\fr{x_{41}^+ x_{32}^+}{x_{31}^+ x_{42}^+},
\end{align}
where the light-cone coordinate $x_j^\pm=t_j \pm x_j$.
In this setup, the light cone limit $z \ll 1-\bar{z} \ll 1$ can be interpreted physically as the limit where the interval $A$ is infinitely boosted; the Cauchy surface containing the intervals becomes singular (see Figure \ref{fig:AB}). That is, the light cone limit corresponds to
\be
x^{-}_{21} \ll x^{+}_{32} \ll 1.
\ee

For simplicity, we assume the interval $B$ to be large ($x_4 \to \infty$ or $x_4 \gg x_3$).
In this case, the cross ratios $z$, $\bar{z}$ are quite simple.
\begin{align}
z=\frac{ x_{21}^-  }{ x_{31} } , \ \
\bar{z} =1- \frac{ x_{32}^+ }{ x_{31} } .
\end{align}
Thus, if we boost $A$ to become almost null, the $n$-th Renyi mutual information (\ref{eq:MI})
can be computed using the four-point function of twist operators
\begin{equation}
\begin{aligned}
 I^{(n)}(A:B) &= \frac{1}{n-1} \log \left[ \fr{ \braket{\sigma_n(x_4) \bar{\sigma}_n(x_3)\sigma_n(x_2) \bar{\sigma}_n(x_1)}  }{\braket{\sigma_n(x_4) \bar{\sigma}_n(x_3)} \braket{\sigma_n(x_2) \bar{\sigma}_n(x_1) }}   \right]\\
&= \frac{1}{n-1} \log \left[ |z|^{4 h_{n}}  G (z, \bar{z}) \right],
\end{aligned}
\end{equation}
where $ G (z, \bar{z}) =\braket{\sigma_n(\infty) \bar{\sigma}_n(1)\sigma_n(z,\bar{z}) \bar{\sigma}_n(0)} $.
The conformal dimension of the twist operator can be written as $h_n =\frac{c}{24} \pa{n-\fr{1}{n}}$; therefore, 
the $n$-th Renyi mutual information is represented as
\begin{equation}\label{eq:generalMI}
\begin{aligned}
 I^{(n)}(A:B) = \frac{c}{12}\pa{1+\fr{1}{n}} \log \fr{x_{31}}{x_{32}^+}   + \fr{1}{n-1}\log \BR{ \abs{z}^{4h_n} \abs{1-z}^{4h_n} G(z, \bar{z})}.
\end{aligned}
\end{equation}

The $n$-th Renyi mutual information in CFTs defined by a complex scalar boson compactified on a torus is calculated in \cite{Caputa2017}.
When the radius of a torus $\eta=\fr{p}{q}$ is rational, 
 \ba
I^{(n)}(A,B)=\frac{c}{12} \pa{1+\frac{1}{n} } \log \pa{ \frac{x_{31}}{x_{32}^+} } -\log(2pq) .
\ea
We expect that, in any rational CFT, this could be generalized into the following form:
\ba
I^{(n)}(A,B)=\frac{c}{12} \pa{1+\frac{1}{n} } \log \pa{ \frac{x_{31}}{x_{32}^+} } -\log d_{tot},
\ea
where $d_{tot}=1/s_{00}$ is the total quantum dimension of the (seed) CFT.
When $\eta$ is irrational,
we obtain the double logarithmic divergent term.
\ba
I^{(n)}(A:B)
= \frac{c}{12} \pa{1+\frac{1}{n} } \log \pa{ \frac{x_{31}}{x_{32}^+} } - \log \pa{ \log \pa{ \frac{x_{31}}{x_{32}^+} } } - \frac{\log n}{n-1} + \log (2\pi).
\ea

In general, the function $G(z, \bar{z})$ is nontrivial, but we can approximate it in the light cone limit as
\footnote{Here, we assume that there are no extra currents, which is expected in generic holographic CFTs. We also assume that orbifoldisation does not change the essential features of the CFTs. It is nontrivial, but actually we can reproduce the holographic results under this assumption; therefore, we expect that this assumption is valid. We will discuss this topic in detail in Sections \ref{subsec:DMI2}.} 
\begin{equation}\label{eq:approximation}
G(z, \bar{z}) \ar{z \ll 1-\bar{z} \ll 1} 
 \ca{F}^{\sigma_n \bar{\sigma}_n}_{ \bar{\sigma}_n \sigma_n}(0|z)\overline{\ca{F}^{\sigma_n \bar{\sigma}_n}_{ \bar{\sigma}_n \sigma_n} (0|\bar{z})},
\end{equation}
where the conformal blocks are defined in a CFT with central charge $nc$. (not $c$ !)
Therefore, we can apply our light cone limit conformal blocks for calculating the $n$-th Renyi mutual information. For simplicity, we first assume a large $c$ limit.
\footnote{
The twist operator has a conformal dimension of the form $c \times const.$. On the other hand, the $c$ dependence of the conformal blocks in the light cone limit appears as $(c-1)\times const.$. Therefore, the blocks with the twist operators have a complicated factor $\fr{c}{c-1}$, which is only simplified in the large $c$ limit. 
}
The vacuum block with twist operators is given as
\begin{equation}
\begin{aligned}
 \ca{F}^{\sigma_n \bar{\sigma}_n}_{ \bar{\sigma}_n \sigma_n}(0|z)\overline{\ca{F}^{\sigma_n \bar{\sigma}_n}_{ \bar{\sigma}_n \sigma_n} (0|\bar{z})}
\ar{z \ll 1-\bar{z} \ll 1} &\left\{
    \begin{array}{ll}
     z^{-2h_n}\pa{1-\bar{z}}^{ -\fr{c}{12n}\pa{1-n}^2 } ,& \text{if} \ \ \ h_n<\fr{nc}{32} , \\
    z^{-2h_n}\pa{1-\bar{z}}^{\fr{c}{24n}\pa{2-n^2} }  ,& \text{ otherwise}.\\
    \end{array}
  \right.\\
\end{aligned}
\end{equation}
 Inserting this vacuum block into the function $G(z,\bar{z})$ in (\ref{eq:generalMI}) leads to the following result:
\begin{equation}\label{eq:MIresult}
\begin{aligned}
 I^{(n)}(A:B)
\ar{z \ll 1-\bar{z} \ll 1} &\left\{
    \begin{array}{ll}
    \frac{c}{12} \pa{1-\frac{1}{n} } \log \pa{ \frac{x_{31}}{x_{32}^+} } ,& \text{if} \ \ \ n<n_* \equiv2 , \\
    \frac{c}{12} \pa{1-\frac{1}{n} } \log \pa{ \frac{x_{31}}{x_{32}^+} }-\fr{c}{24}\fr{(n-2)^2}{n(n-1)}\log \pa{ \frac{x_{31}}{x_{32}^+}}   ,& \text{ otherwise}.\\
    \end{array}
  \right.\\
\end{aligned}
\end{equation}
In particular, the mutual information is given by taking the limit $n \to 1$ as
\ba
I(A:B) \ar{n \to 1} 0.
\ea
This result is consistent with the holographic calculation as in \cite{Kusuki2017}. We intend to emphasize that 
the additional logarithmic divergent term appears in the $n$-th Renyi mutual information for $n>n_*$. 
In many cases, to calculate the entanglement entropy (for example, the replica method), we implicitly assume that the Renyi entropy is analytic in $n$. However, we find an exception of this assumption in the light cone limit. Therefore, we have to consider this exception if we use the replica method to evaluate the entanglement entropy. We emphasize that this assumption does not contradict with the derivation of the Ryu--Takayanagi formula in \cite{Lewkowycz2013}, as our result for the Renyi entropy is analytic in the vicinity of $n=1$.

\begin{figure}[t]
 \begin{center}
  \includegraphics[width=8.0cm,clip]{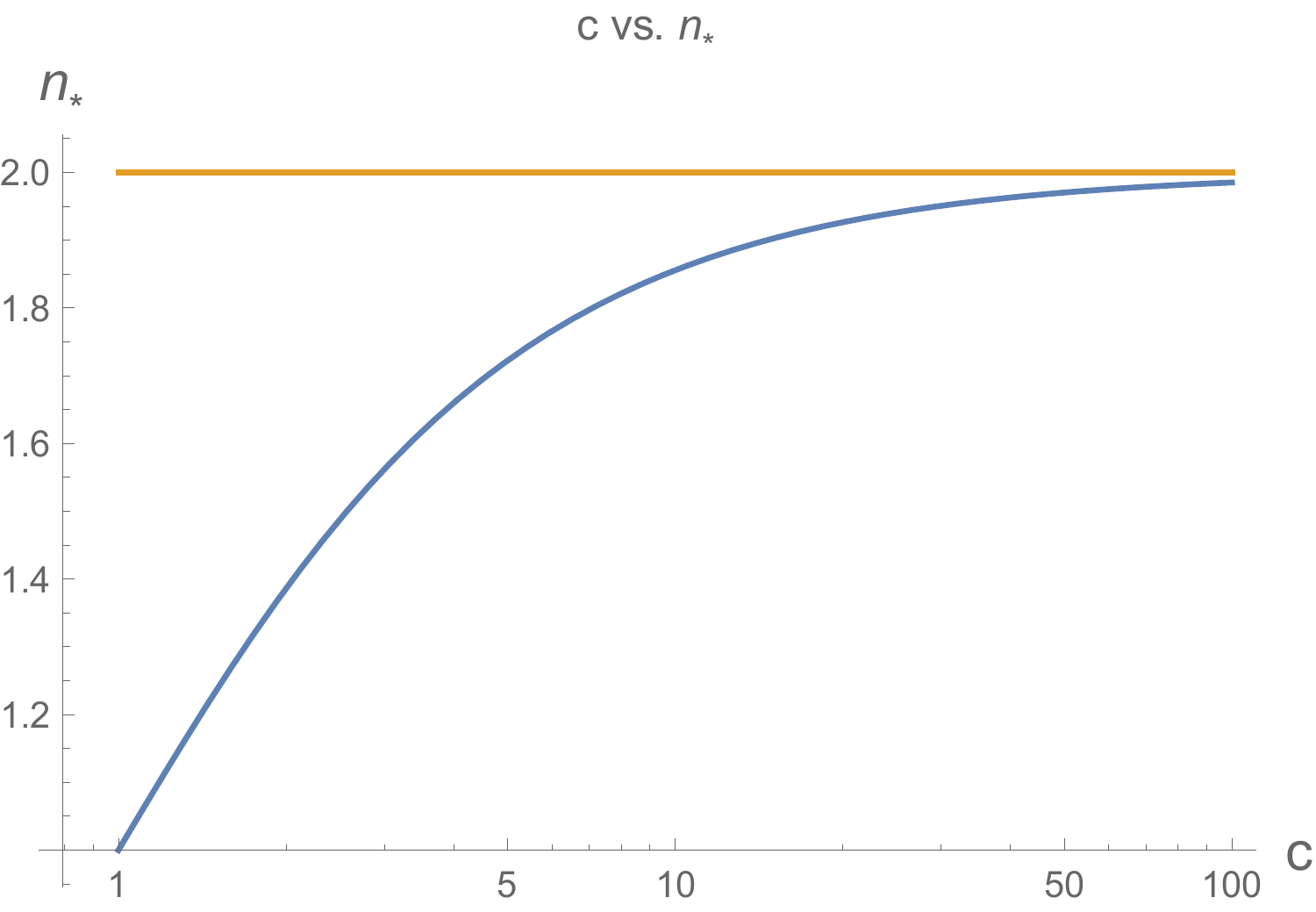}
 \end{center}
 \caption{$c$ dependence of $n_*$. The classical limit $c \to \infty$ matches $n_*=2$.}
 \label{fig:point}
\end{figure}

 We expect that this phase transition arises from only the light cone limit $z \ll 1-\bar{z} \ll 1$ (and $c>1$) and not the large $c$ limit. 
Following (\ref{eq:FMresult1}) (or (\ref{eq:FMresult2})) in Appendix \ref{subapp:FM}, we can immediately obtain the Renyi mutual information for CFTs with finite $c$ as 
\begin{equation}
 I^{(n)}(A:B)
\ar{z \ll 1-\bar{z} \ll 1}
\fr{c}{12}\pa{\pa{1+\fr{1}{n}}-\fr{s_n}{n-1}} \log \pa{ \frac{x_{31}}{x_{32}^+} },
\end{equation}
and the function $s_n$ is given by
\begin{equation}
\begin{aligned}
s_n& = \left\{
    \begin{array}{ll}
    2\a_n(Q-2\a_n)  ,& \text{if } 2\a_n<\fr{Q}{2}\   ,\\
    \fr{Q^2}{4} ,& \text{otherwise }  ,\\
    \end{array}
  \right.\\
\end{aligned}
\end{equation}
where $nc=1+6Q^2$ and  $\a_n=\fr{Q}{2}\pa{1-\s{1-\fr{c}{nc-1}\pa{n-\fr{1}{n}}}}$, which satisfies $h_n=\a_n(Q-\a_n)$.
The transition point for general $c$ is given by a more complicated form than $n_*=2$ in (\ref{eq:DMI}) as follows:
\begin{equation}
n_*=\fr{3}{2c} \pa{\s{1+\fr{16}{9}c^2}-1},
\end{equation}
which satisfies $n_*\ar{c \to \infty} 2 $ as expected. The $c$ dependence of $n_*$ is shown in Figure \ref{fig:point}.
This shows that the transition point is located between $1<n_*<2$. Therefore, deduction of entanglement from the Renyi mutual information has to be done carefully.

In particular, we obtain the limit $n \to 1$ of the Renyi mutual information as
\begin{equation}
I^{(n)}(A:B)\ar{n \to 1} \fr{c^2}{12} \fr{n-1}{c-1}.
\end{equation}
This suggests that in CFTs with $c>1$ (and no conserved primary currents), the mutual information in the light cone limit vanishes as in holographic CFTs. On the other hand, if we consider a CFT with $c=1$, the mutual information becomes ill-defined.
This is natural because the equation (\ref{eq:FMresult1}) (or (\ref{eq:FMresult2})) is valid only if $c>1$, as mentioned in Appendix \ref{subapp:FM}.
We attribute this to the same reason why the quasiparticle picture breaks down if we assume no extended symmetry algebra and $c>1$, as explained in \cite{Asplund2015a} (see also Section \ref{subsec:DMI}, \ref{subsec:DMI2}).
We can also find similar Renyi phase transitions as the replica number $n$ varies in other situations \cite{Metlitski2009,Belin2013,Belin2015,Belin2017,Dong2018}.

\begin{figure}[t]
 \begin{center}
  \includegraphics[width=7.0cm,clip]{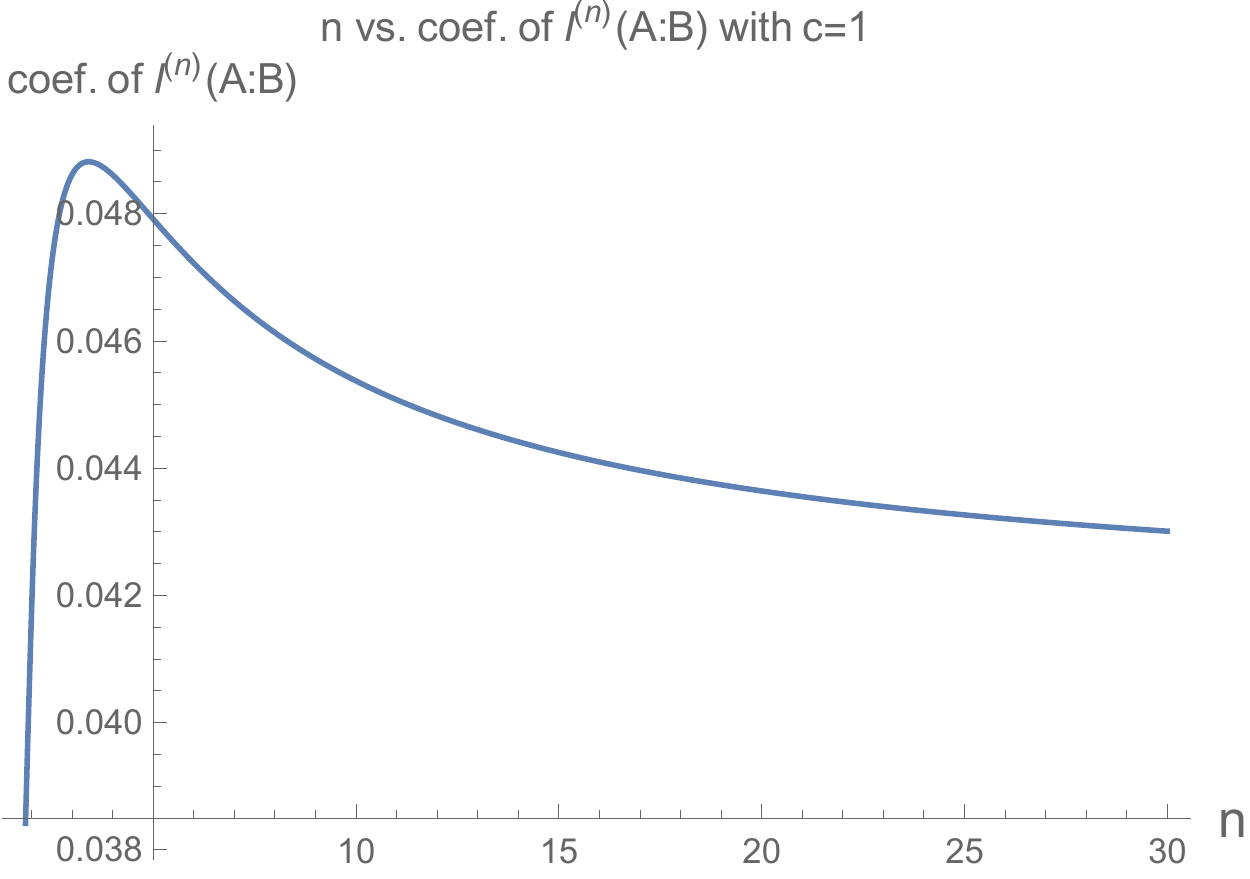}
  \includegraphics[width=7.0cm,clip]{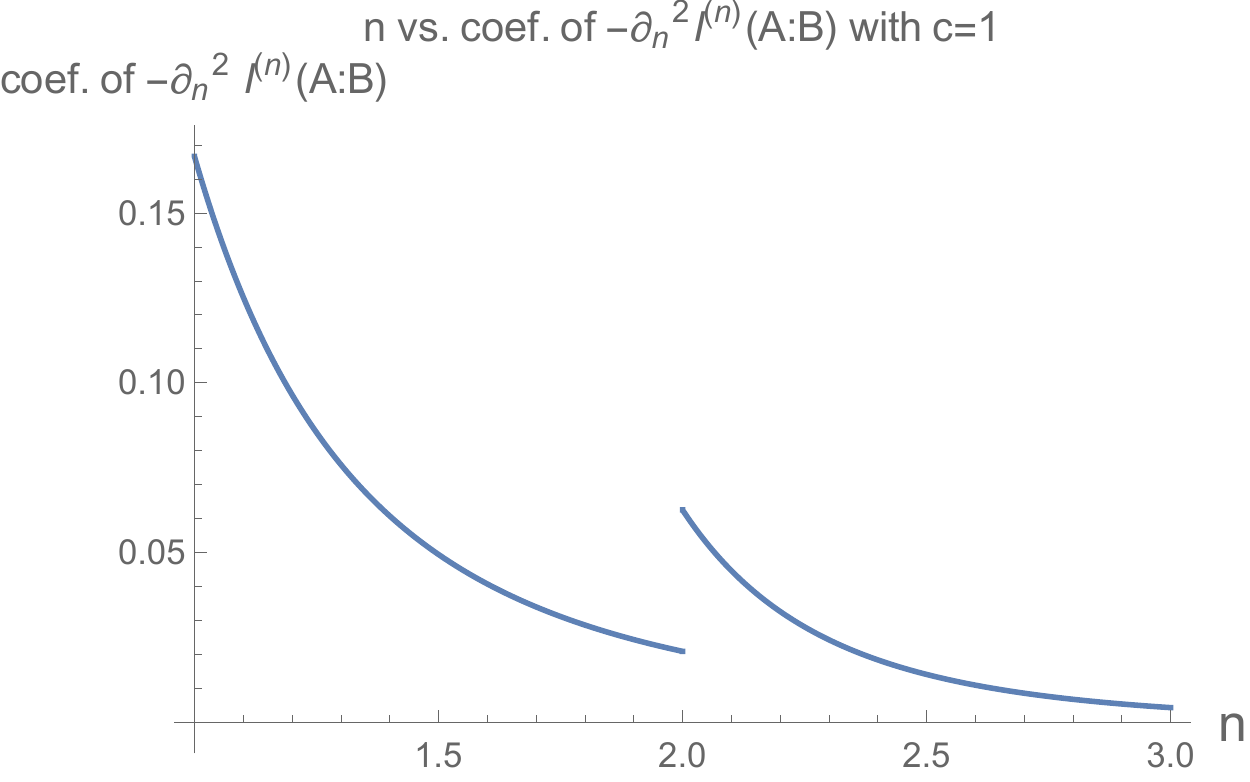}
 \end{center}
 \caption{The left figure shows the $n$ dependence of the coefficient of the Renyi mutual information (\ref{eq:MIresult}). It can be observed that the $n \to \infty$ limit of the Renyi mutual information approaches $I^{(2)}(A:B)$. The right figure shows the $n$ dependence of the coefficient of $\del^2_n I^{(n)}(A:B)$. One can see that $\del^2_n I^{(n)}(A:B)$ becomes discontinuous at $n=n_*$.}
 \label{fig:MIplot}
\end{figure}

It would be interesting to point out that (\ref{eq:MIresult}) leads to the fact that the $(n>2)$-th Renyi mutual information in the light cone limit is bounded by the 2nd Renyi mutual information as follows:
\begin{equation}
 I^{(2)}(A:B)<I^{(n)}(A:B) \ \ \ \ \text{if} \ \ n>2,
\end{equation}
and in particular,
\footnote{$S^{(2)}$ and $S^{(\infty)}$ are respectively known as {\it collision entropy} and {\it min-entropy}. }
\begin{equation}
 \lim_{n \to \infty}  I^{(n)}(A:B) = I^{(2)}(A:B).
\end{equation}
These properties are depicted in Figure \ref{fig:MIplot}.

\subsection{Dynamics of Renyi Mutual Information in Large $c$ CFTs}\label{subsec:DMI}

The dynamics of quantum information has attracted the attention of many research communities. In this sense, we are also interested in the propagation of entanglement. Fortunately, as discussed in \cite{Asplund2015a}, entanglement memory could be characterised using light cone singularity; therefore, we might be able to study entanglement memory using our conformal blocks in the light cone limit. In this section, we explain this in detail and apply our conformal blocks to the calculation of the entanglement entropy.

\begin{figure}[t]
 \begin{center}
  \includegraphics[width=7.0cm,clip]{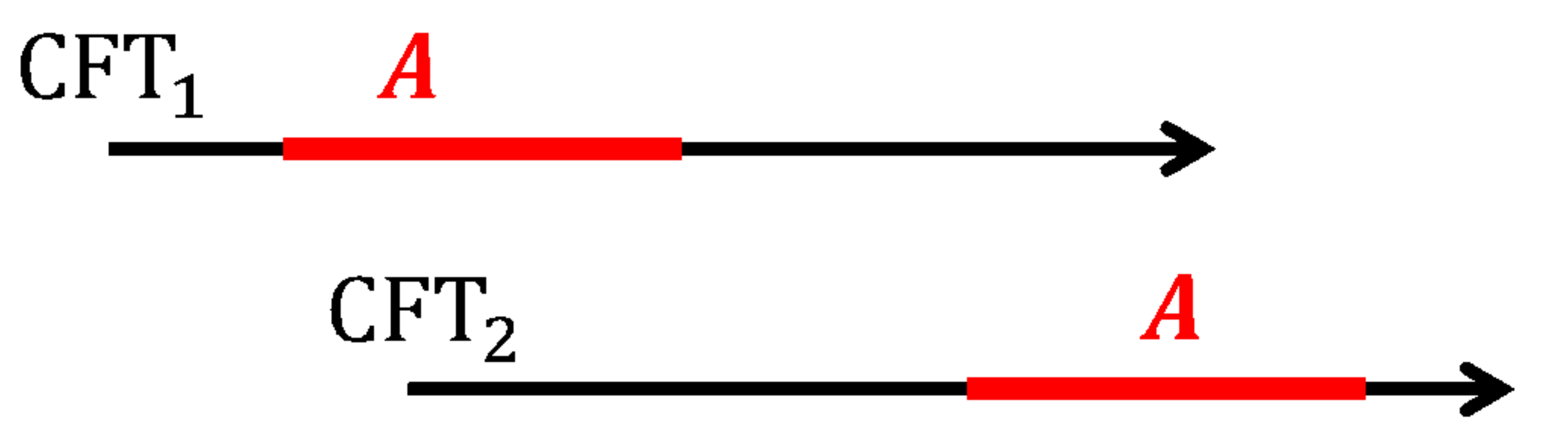}
  \includegraphics[width=7.0cm,clip]{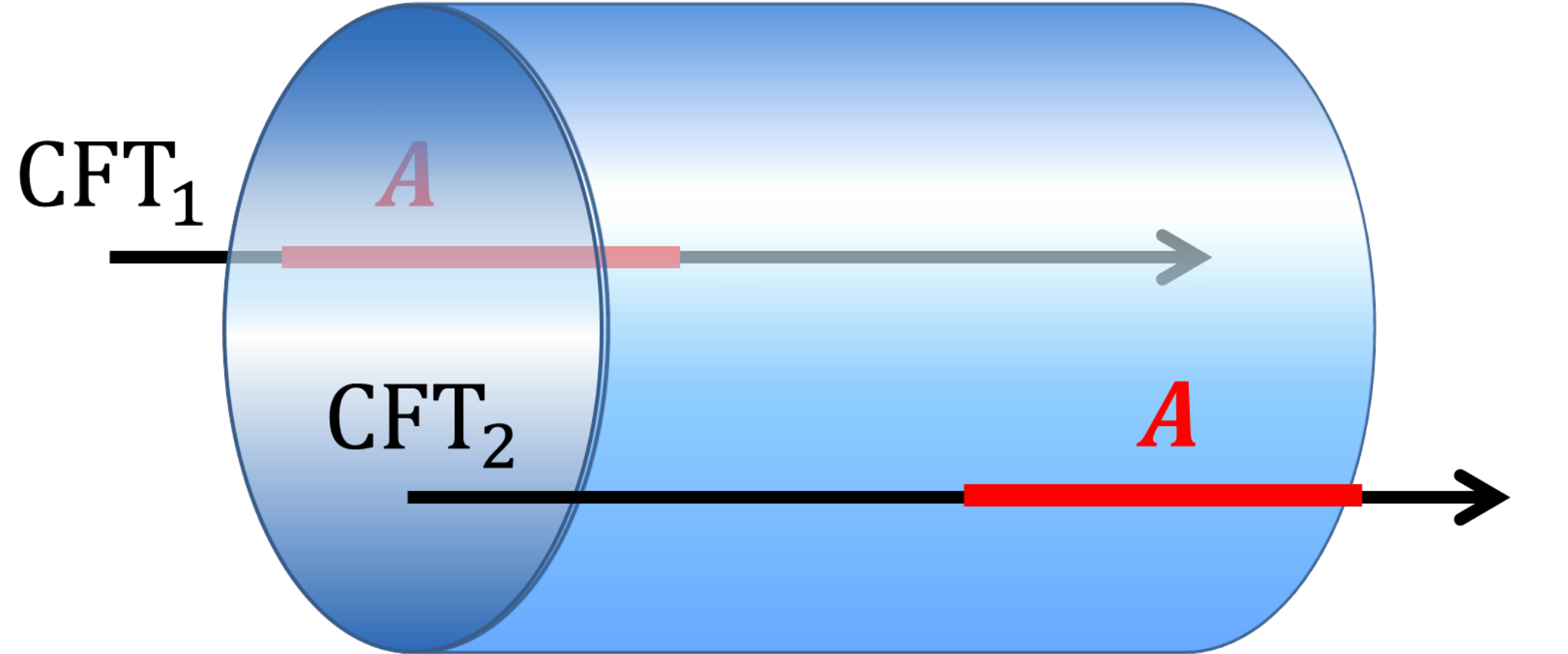}
 \end{center}
 \caption{The left figure shows two intervals in the doubled CFT. The entanglement entropy in this setup can be given by a four-point function with twist operators on a thermal cylinder of periodicity $\beta$, as sketched in the right figure.}
 \label{fig:TFD}
\end{figure}

For realizing the objective stated above, we consider the Renyi entropy in a doubled CFT. That is, we will consider the thermofield double state in the doubled system,
\begin{equation}
\ket{\text{TFD}}=\sum_n\ex{-\fr{\b}{2} H} \ket{n}_1 \ket{n}_2,
\end{equation}
as an entangled state. Two states $\ket{n}_1$ and $\ket{n}_2$ respectively exist in CFT${}_1$ and CFT${}_2$. We label the coordinates of each CFT as $(t_1,x_1)$ and $(t_2,x_2)$ and consider two disconnected intervals (see the left of Figure \ref{fig:TFD}) 
\begin{equation}
A=[0,L]_1 \cup[D+L,D+2L]_2.
\end{equation}
Let us choose the total Hamiltonian acting on the doubled CFT as
\begin{equation}
H_{\text{tot}}=H_1+H_2.
\end{equation}
Thus, the thermofield-double state (TFD) has a non-trivial time dependence. 

The Renyi entropy in a doubled CFT can be given by a four-point function with twist operators on a thermal cylinder of periodicity $\beta$. The twist operators are put on the endpoints of $A$ with a shift $i\fr{\beta}{2}$ for operators in two different copies of the CFT. The time dependence is obtained by considering the analytic continuation $t \to it$ of the insertion points of the twist operators (see the right of Figure \ref{fig:TFD}),

\begin{equation}
\pa{\fr{2\pi}{\beta}}^{8h_n} \abs{w_1 w_2 w_3 w_4}^{2h_n}\braket{\sigma_n(w_1,\bar{w}_1) \bar{\sigma}_n(w_2,\bar{w}_2) \sigma_n(w_3,\bar{w}_3) \bar{\sigma}_n(w_4,\bar{w}_4)  },
\end{equation}
where the insertion points are given by
\begin{align}
w_1 &= \ex{\frac{2\pi}{\beta}(-t+i\beta/4)} ,  &  \bw_1 &= \ex{\frac{2\pi}{\beta}(t-i\beta/4)},\notag\\
w_2 &= \ex{\frac{2\pi}{\beta}(L-t+i\beta/4)} , & \bw_2 &= \ex{\frac{2\pi}{\beta}(L + t - i\beta/4)},\\
w_3 &= \ex{\frac{2\pi}{\beta}(D+2L+t-i\beta/4)} , & \bw_3 &= \ex{\frac{2\pi}{\beta}(D+2L-t+i\beta/4)},\notag\\
w_4 &= \ex{\frac{2\pi}{\beta}(D+L+t-i\beta/4)} , & \bw_4 &= \ex{\frac{2\pi}{\beta}(D+L-t+i\beta/4)}\notag \ .
\end{align}
The Renyi mutual information can be obtained in a simpler manner as follows:
\begin{equation}
\begin{aligned}
 I^{(n)}_A &= \frac{1}{n-1} \log \left[ \fr{\braket{\sigma_n(w_1,\bar{w}_1) \bar{\sigma}_n(w_2,\bar{w}_2) \sigma_n(w_3,\bar{w}_3) \bar{\sigma}_n(w_4,\bar{w}_4)  }  }{\braket{\sigma_n(w_1,\bar{w}_1) \bar{\sigma}_n(w_2,\bar{w}_2)}\braket{ \sigma_n(w_3,\bar{w}_3) \bar{\sigma}_n(w_4,\bar{w}_4)  }}   \right]\\
&= \frac{1}{n-1} \log \left[ |z|^{4 h_{n}}  G (z, \bar{z}) \right],
\end{aligned}
\end{equation}
where $ G (z, \bar{z}) =\braket{\sigma_n(\infty) \bar{\sigma}_n(1)\sigma_n(z,\bar{z}) \bar{\sigma}_n(0)} $ and $z=\fr{w_{12}w_{34}}{w_{13} w_{24}}$.

In general, a four-point function non-trivially depends on the CFT data, and its explicit form is not known, except for a few CFTs. Therefore, to proceed further, we focus on the high-temperature limit $\b \to 0$ in the following, which simplifies the four-point function as it corresponds to some OPE limits. That is, the cross ratio is given by
\begin{equation}
\begin{aligned}
z &\simeq \ex{-\fr{2\pi}{\b}(D+2t)} \ar{\b \to 0} 0,\\
\bar{z} &\simeq \ex{-\fr{2\pi}{\b} \pa{ D+2L+t -\text{max}(D+2L-t,t)-\text{max}(D,2t) }} \ar{\b \to 0} 0 \  \text{or} \  1.
\end{aligned}
\end{equation}
Outside the range $\fr{D}{2}<t<\fr{D+2L}{2}$, the anti-holomorphic cross ratio approaches $0$; thus, the function $G(z,\bar{z})$ becomes
\begin{equation}
G(z,\bar{z}) \ar{\b \to 0} \abs{z}^{-4h_n}.
\end{equation}
As a result, we obtain the Renyi mutual information as
\begin{equation}
I_A \ar{\b \to 0} 0.
\end{equation}
On the other hand, in the range $\fr{D}{2}<t<\fr{D+2L}{2}$, the anti-holomorphic cross ratio is given by
\begin{equation}
1-\bar{z}\simeq \ex{-\fr{2\pi}{\b}\text{min}(D+2L-2t,2t-D)} \ar{\b\to0}0.
\end{equation}
In particular, this cross ratio satisfies $z \ll 1-\bar{z} \ll 1$, and therefore, this limit corresponds to the light cone limit. Thus, the calculation of the Renyi mutual information in the doubled CFT reduces to the same form as given in Section \ref{subsec:MILC}.
The function $G(z,\bar{z})$ is approximated as
\begin{equation}
\begin{aligned}
G(z,\bar{z})
\ar{z \ll 1-\bar{z} \ll 1} &\left\{
    \begin{array}{ll}
     z^{-2h_n}\pa{1-\bar{z}}^{ -\fr{c}{12n}\pa{1-n}^2 } ,& \text{if} \ \ \ h_n<\fr{nc}{32} , \\
    z^{-2h_n}\pa{1-\bar{z}}^{\fr{c}{24n}\pa{2-n^2} }  ,& \text{ otherwise},\\
    \end{array}
  \right.\\
\end{aligned}
\end{equation}
and, therefore, the Renyi mutual information is
\begin{equation}\label{eq:DMI}
\begin{aligned}
 I^{(n)}_A
\ar{z \ll 1-\bar{z} \ll 1} &\left\{
    \begin{array}{ll}
    \frac{c}{12} \pa{1-\frac{1}{n} } \fr{2\pi}{\b}\text{min}(D+2L-2t,2t-D) ,& \text{if} \ \ \ n<n_* \equiv2 , \\
    \BR{\frac{c}{12} \pa{1-\frac{1}{n} } -\fr{c}{24}\fr{(n-2)^2}{n(n-1)}}\fr{2\pi}{\b}\text{min}(D+2L-2t,2t-D)   ,& \text{ otherwise}.\\
    \end{array}
  \right.\\
\end{aligned}
\end{equation}
As a result, one can again see the Renyi phase transition as the replica number is varied. Therefore, one has to take care when trying to predict the behaviour of the entanglement entropy using the Renyi entropy.

We have to mention that from (\ref{eq:DMI}), the mutual information is obtained by taking the limit $n \to 1$ as
\begin{equation}
I_A=0.
\end{equation}
Here, the mutual information vanishes for all times. It means that entanglement scrambles maximally, which contradicts with the quasiparticle behaviour shown in, for example, rational CFTs.

\subsection{Renyi Mutual Information beyond Large $c$}\label{subsec:DMI2}

As mentioned in Section \ref{subsec:MILC}, we can generalize the calculation of the Renyi mutual information to general $c$ under some assumptions. This is extensively discussed in this section. 

For simplicity, we again assume that there are no extra currents. Even in such a case, orbifoldisation leads to the $\bb{Z}_n$ current; therefore,  we should approximate a correlator with twist operators, instead of (\ref{eq:approximation}), as follows:
\begin{equation}
G(z, \bar{z}) \ar{z \ll 1-\bar{z} \ll 1} 
 \ca{F}^{\sigma_n \bar{\sigma}_n}_{ \bar{\sigma}_n \sigma_n}(0|z)\overline{{\ca{F}^{\text{Vir}^n/\bb{Z}_n}}^{\sigma_n \bar{\sigma}_n}_{ \bar{\sigma}_n \sigma_n} (0|\bar{z})},
\end{equation}
where $\ca{F}^{\text{Vir}^n/\bb{Z}_n}$ is the conformal block defined by current algebra $\text{Vir}^n/\bb{Z}_n$ and not just Virasoro algebra.

From the crossing symmetry, we can obtain 
\begin{equation}
G(z,\bar{z})=G(1-z,1-\bar{z})\ar{\bar{z} \to 1} (1-\bar{z})^{-2h_n}.
\end{equation}
Therefore, we have the upper bound of the singularity of $\ca{F}^{\text{Vir}^n/\bb{Z}_n}$ as
\begin{equation}
\lim_{\bar{z}\to1}\overline{{\ca{F}^{\text{Vir}^n/\bb{Z}_n}}^{\sigma_n \bar{\sigma}_n}_{ \bar{\sigma}_n \sigma_n} (0|\bar{z})} \lsim(1-\bar{z})^{-2h_n},
\end{equation}
In addition, we can also deduce the lower bound as
\begin{equation}
\begin{aligned}
\lim_{\bar{z}\to1}\overline{{\ca{F}^{\text{Vir}^n/\bb{Z}_n}}^{\sigma_n \bar{\sigma}_n}_{ \bar{\sigma}_n \sigma_n} (0|\bar{z})} 
\gsim \lim_{\bar{z}\to1}\overline{\ca{F}^{\sigma_n \bar{\sigma}_n}_{ \bar{\sigma}_n \sigma_n} (0|\bar{z})}.
\end{aligned}
\end{equation}
The light cone singularity of the Virasoro block with any $c>1$ is given in Appendix \ref{subapp:FM},
\begin{equation}
\begin{aligned}
\ca{F}^{AA}_{BB}(h_{\a_s}|z)& \ar{z\to 1}  (1-\bar{z})^{s_n-2h_n},
\end{aligned}
\end{equation}
where the function $s_n$ is defined as
\begin{equation}
\begin{aligned}
s_n& = \left\{
    \begin{array}{ll}
    2\a_n(Q-2\a_n)  ,& \text{if } 2\a_n<\fr{Q}{2}\   ,\\
    \fr{Q^2}{4} ,& \text{otherwise }  .\\
    \end{array}
  \right.\\
\end{aligned}
\end{equation}
Here, $nc=1+6Q^2$ and  $\a_n=\fr{Q}{2}\pa{1-\s{1-\fr{c}{nc-1}\pa{n-\fr{1}{n}}}}$, which satisfies $h_n=\a_n(Q-\a_n)$.
In conclusion, the light cone limit of a correlator $G(z,\bar{z})$ is bounded by
\begin{equation}
z^{-2h_n} (1-\bar{z})^{s_n-2h_n}
\lsim G(z,\bar{z})
\lsim z^{-2h_n} (1-\bar{z})^{-2h_n}.
\end{equation}

As a result, the Renyi mutual information in the doubled CFTs satisfies the following inequalities:
\begin{equation}\label{eq:Ibound}
\begin{aligned}
 \pa{\frac{c}{12} \pa{1+\frac{1}{n} }-\fr{s_n}{n-1}} \fr{2\pi}{\b}\text{min}(D+2L-2t,2t-D)
&\leq I^{(n)}(A:B)\\
&\leq \frac{c}{12} \pa{1+\frac{1}{n} } \fr{2\pi}{\b}\text{min}(D+2L-2t,2t-D).
\end{aligned}
\end{equation}
In fact, rational CFTs saturate the upper bound (\ref{eq:Ibound}), and hence, their mutual information is universal.
This shows that entanglement does not scramble and quasiparticle behaviour can be observed in rational CFTs.
On the other hand, holographic CFTs appear to saturate the lower bound. In this context, we can state that if the Renyi mutual information, in theory, saturates the lower bound, then it shows maximal scrambling. This is the main conclusion of this section.

\subsection{2nd Renyi Entropy after Local Quench}\label{subsec:2ndREE}

\begin{figure}[t]
 \begin{center}
  \includegraphics[width=5.0cm]{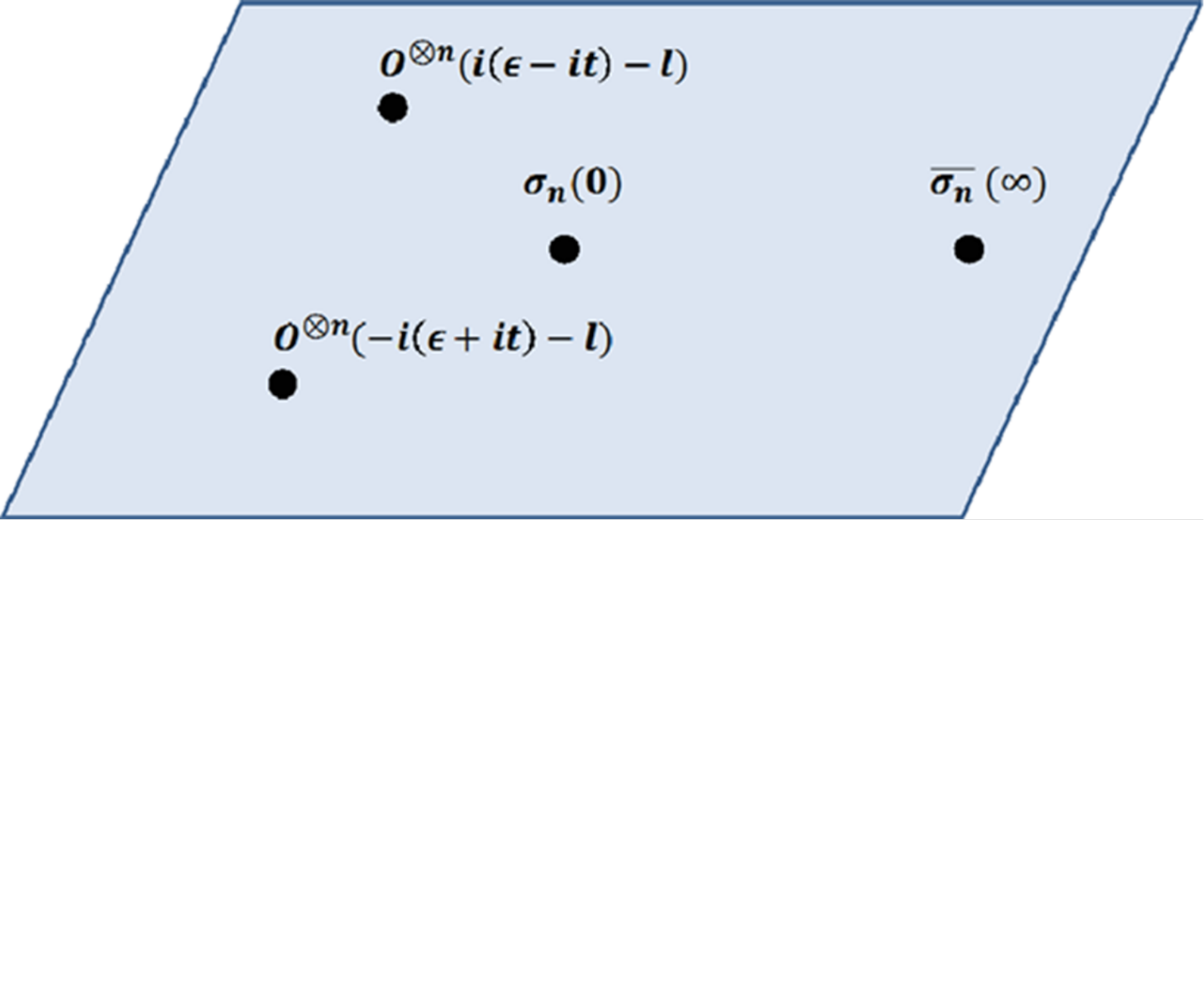}
  \includegraphics[width=9.0cm]{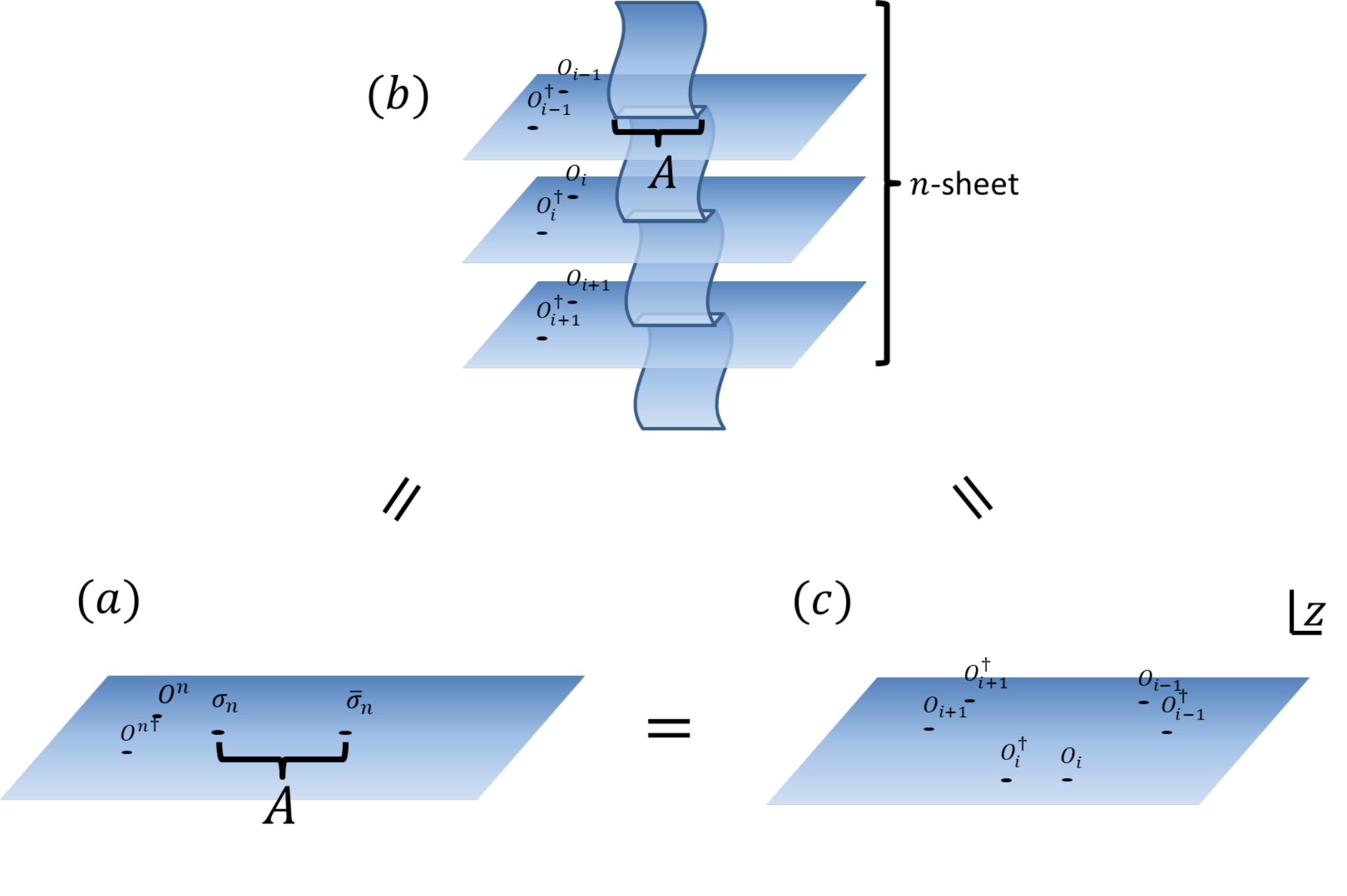}
 \end{center}
 \caption{(Left) The positions of operators in the replica computation (\ref{eq:defREE}). (Right) The equivalence between $(a)$ and $(b)$ explains the relation between a correlator with twist operators in an orbifold theory and a replica manifold. The equivalence between $(b)$ and $(c)$ can be obtained using a conformal map $w=z^n$.}
 \label{fig:pos}
\end{figure}

At the end of this section, we discuss another application of light cone singularity.
In fact, the light cone limit also appears if one investigates the dynamics of the Renyi entanglement entropy after a local quench. The process is as follows: We consider the locally excited state $\ket{\Psi}$, which is defined by acting with a local operator $O(x)$ on the CFT vacuum $\ket{0}$ in the following manner,\footnote{We would like to stress that
$\ep$ in (\ref{lopw}) is the ultraviolet (UV) cut off of the local excitations and should be distinguished from the UV cut off (i.e. the lattice spacing) of the CFT itself.}
\be\label{lopw}
\ket{\Psi(t)}=\ca{N}\ex{-\ep H-iHt}O(x)\ket{0}, 
\ee
where $\ca{N}$ is the normalization factor. The infinitesimally small parameter $\ep>0$ provides UV regularization as the truly localized operator has infinite energy. We choose the subsystem $A$ to be the half-space and induce excitation in its complement, thus creating additional entanglements between them. The main quantity of interest is the growth of entanglement entropy compared to the vacuum:
 \be\label{difs}
 \Delta S^{(n)}_A(t)=S^{(n)}_A(\ket{\Psi(t)})-S^{(n)}_A(\ket{0}).
 \ee
In fact, this quantity can also be calculated analytically using twist operators \cite{Nozaki2014} as
\begin{equation}\label{eq:defREE}
\Delta S^{(n)}_A=\frac{1}{1-n}\log \frac{\ave{O^{\otimes n}O^{\otimes n}\sigma_n \bar{\sigma_n} }}{\ave{O^{\otimes n}O^{\otimes n}}\ave{\sigma_n \bar{\sigma_n}}},
\end{equation}
where the operator $O^{\otimes n}$ is defined on the cyclic orbifold CFT $\ca{M}^n/\bb{Z}_n$, using the operators in the seed CFT $\ca{M}$ as
\begin{equation}
O^{\otimes n} = O \otimes O \otimes \cdots \otimes O.
\end{equation}
The local excitation $O$ is separated by a distance $l$ from the boundary of $A$, as shown in the left of Figure \ref{fig:pos}.
For simplicity, we move the framework from $(a)$ to $(c)$ (through $(b)$) as shown in the right of Figure \ref{fig:pos} using a conformal map $w=z^n$. We focus on $n=2$ in the following. In the late-time limit $t \gg l$, we can approximate the cross ratio as
\begin{equation}
z\simeq1-\fr{\e^2}{4t^2}, \ \ \ \ \ \bar{z}\simeq\fr{\e^2}{4t^2},
\end{equation}
which is just the light cone limit. \footnote{Actually, this limit is the double light cone limit, which is defined by the limit $1-z,\bar{z} \ll 1$ with $\fr{\bar{z}}{1-z}$ fixed.}

Using this cross ratio, we can re-express the correlator as
\begin{equation}
 \frac{\ave{O^{\otimes n}O^{\otimes n}\sigma_n \bar{\sigma_n} }}{\ave{O^{\otimes n}O^{\otimes n}}\ave{\sigma_n \bar{\sigma_n}}}
=\abs{z}^{4h_O}\abs{1-z}^{4h_O}G(z,\bar{z}).
\end{equation}
The light cone limit of the four-point function can be approximated as
\footnote{In this case, the function $G(z,\bar{z})$ is NOT defined in an orbifold theory, but just in an ordinal theory. Therefore, we do not need to consider the difficulty explained in Section \ref{subsec:DMI2}.  We can approximate the light cone limit of a correlator using just the Virasoro conformal block.}
\begin{equation}
G(z,\bar{z})\ar{1-z,\bar{z} \ll 1}  (1-z)^{s_O-2h_O}\bar{z}^{ -2h_O} ,
\end{equation}
where the function $s_O$ is defined as 
\begin{equation}
\begin{aligned}
s_O& = \left\{
    \begin{array}{ll}
    2\a_O(Q-2\a_O)  ,& \text{if } h_O<\fr{c-1}{32}\   ,\\
    \fr{Q^2}{4} ,& \text{otherwise }  .\\
    \end{array}
  \right.\\
\end{aligned}
\end{equation}
Therefore, the growth of the $2$nd Renyi entropy after a light local quench ($h_O\leq\fr{c-1}{32}$)  is given by
\begin{equation}
\D S_A^{(2)}(t)\ar{\fr{t}{\e} \to \infty}4\a_O(Q-2\a_O)\log\fr{t}{\e}.
\end{equation}
In particular, if expanding this at small $\fr{h_O}{c}$, the result reduces to
\begin{equation}
\D S_A^{(2)}\ar{\fr{h_O}{c}\ll1}4h_O\log\fr{t}{\e}.
\end{equation}
This result in the light limit is consistent with the result in \cite{Caputa2014a}.
The growth for a heavy local quench ($h_O\geq\fr{c-1}{32}$) is more interesting, that is, it has the following universal form:
\begin{equation}\label{eq:univREE}
\D S_A^{(2)}(t)\ar{\fr{t}{\e} \to \infty}\fr{Q^2}{2} \log\fr{t}{\e}.
\end{equation}
These results are consistent with the numerical results in \cite{Kusuki2018b}.

\begin{figure}[t]
 \begin{center}
  \includegraphics[width=10.0cm]{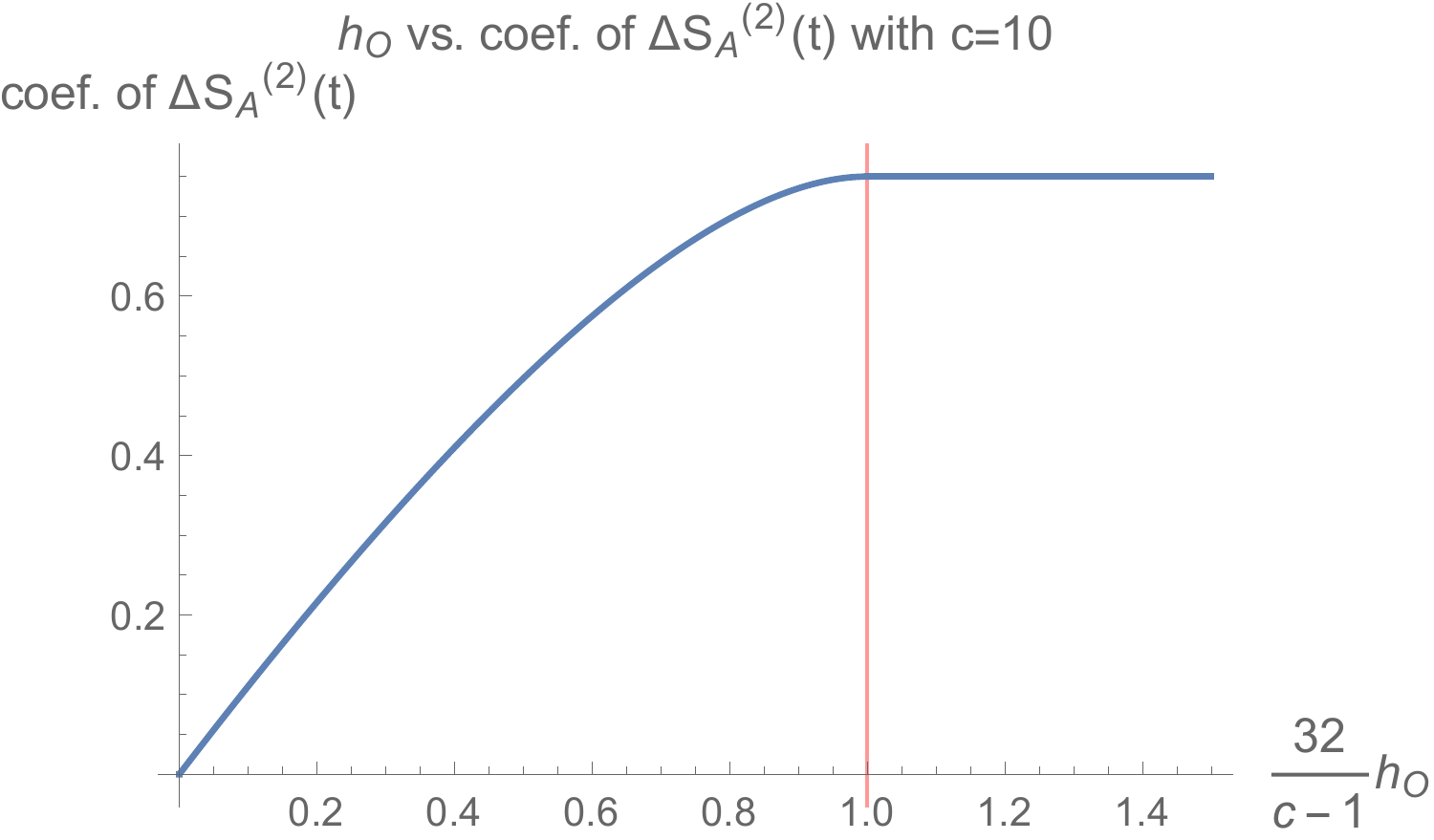}
 \end{center}
 \caption{The $h_O$ dependence of the coefficient of the growth of the $2$nd Renyi entropy after a local quench. This dependence might imply that in a CFT with $c>1$ and no extra currents, the heavier the operator used to create a local quench, the larger is the entropy growth; however, if its dimension exceeds the value $\fr{c-1}{32}$, then the entropy is saturated by (\ref{eq:univREE})}
 \label{fig:2ndREE}
\end{figure}

We can, therefore, conclude that the $2$nd Renyi entropy after a local quench undergoes a phase transition as the conformal dimension of the local quench is varied, if we restrict ourselves to a unitary (compact) CFT with $c>1$ and no extra conserved currents. That is, in one of the phases, the entropy is monotonically increasing in $h_O$, and in the other phase, it is saturated by the universal form (\ref{eq:univREE}), as shown in Figure \ref{fig:2ndREE}. We intend to emphasize that at least when $n=2$, the Renyi entropy after a local quench can be explicitly given without other assumptions except that $c>1$ and there are no extra currents (and discrete spectra or, equivalently, compactness).
Unfortunately, we did not find this saturation when studying the entanglement entropy ($n=1$ Renyi entropy); therefore, we could not determine how to relate this saturation to the dynamics of the entanglement.
Note that, in fact, we can generalize this result to any replica number $n$, as explained in the next section.

\section{Regge Limit Universality}\label{sec:Regge}

In 2D CFTs, the Regge limit is defined by the limit $z, \bar{z}\to 0$ after picking up a monodromy around $z=1$ as $(1-z) \to \ex{-2\pi i}(1-z)$. This limit is obviously different from the light cone limit; nevertheless, we can contribute to studies on the Regge limit using our light cone limit conformal blocks.

For this purpose, we introduce the elliptic form of the Virasoro blocks as follows:
\begin{equation}
\ca{F}^{21}_{34}(h_p|z)=\Lambda^{21}_{34}(h_p|q)H^{21}_{34}(h_p|q),\ \ \ \ \ \ q(z)=\ex{-\pi \fr{K(1-z)}{K(z)}},
\end{equation}
where $K(z)$ is the elliptic integral of the first kind and the function $\Lambda^{21}_{34}(h_p|q)$ is a universal prefactor given by
\begin{equation}\label{eq:preF}
 \Lambda^{21}_{34}(h_p|q)=(16q)^{h_p-\frac{c-1}{24}}z^{\frac{c-1}{24}-h_1-h_2}(1-z)^{\frac{c-1}{24}-h_2-h_3}
(\theta_3(q))^{\frac{c-1}{2}-4(h_1+h_2+h_3+h_4)}.
\end{equation}
The function $H^{21}_{34}(h_p|q)$ can be calculated recursively (see Appendix \ref{app:recursion}).
For simplicity, we express the function $H^{21}_{34}(h_p|q)$ using a series expansion form as
\begin{equation}
H^{21}_{34}(h_p|q)=\sum_{n\in \bb{Z}_{\geq0}} c_n q^{n},
\end{equation}
where $c_0=1$. In our recent studies \cite{Kusuki2018,Kusuki2018b}, we determined that the series coefficients could be expressed as
\begin{equation}
c_n \sim\xi^n  n^{\a} \ex{A\s{n}} \ \ \ \ \ \ \ \ \text{for large $n\gg c$},
\end{equation}
where 
\begin{equation}\label{eq:signxi}
\begin{aligned}
\xi&=\left\{
    \begin{array}{ll}
      \d_{n,\text{even}} \times \sgn\BR{\pa{h_A-\fr{c-1}{32}}\pa{h_B-\fr{c-1}{32}}} ,& \text{for AABB blocks }   ,\\
      1 ,& \text{for ABBA blocks }.\\
    \end{array}
  \right.\\
\end{aligned}
\end{equation}
The values of $A$ and $\a$ are constants in $n$, depending on $h_A, h_B$, and $c$ (whose explicit forms are given in \cite{Kusuki2018,Kusuki2018b,Kusuki2018a} or (\ref{eq:AABBA}) (\ref{eq:AABBal}), (\ref{eq:ABBAA}), (\ref{eq:ABBAal}) in Appendix \ref{app:recursion}). Note that the series coefficients for the AABB blocks vanish if $n$ is odd, which is described by $\d_{n,\text{even}}$ in (\ref{eq:signxi}).

\begin{figure}[t]
 \begin{center}
  \includegraphics[width=12.0cm,clip]{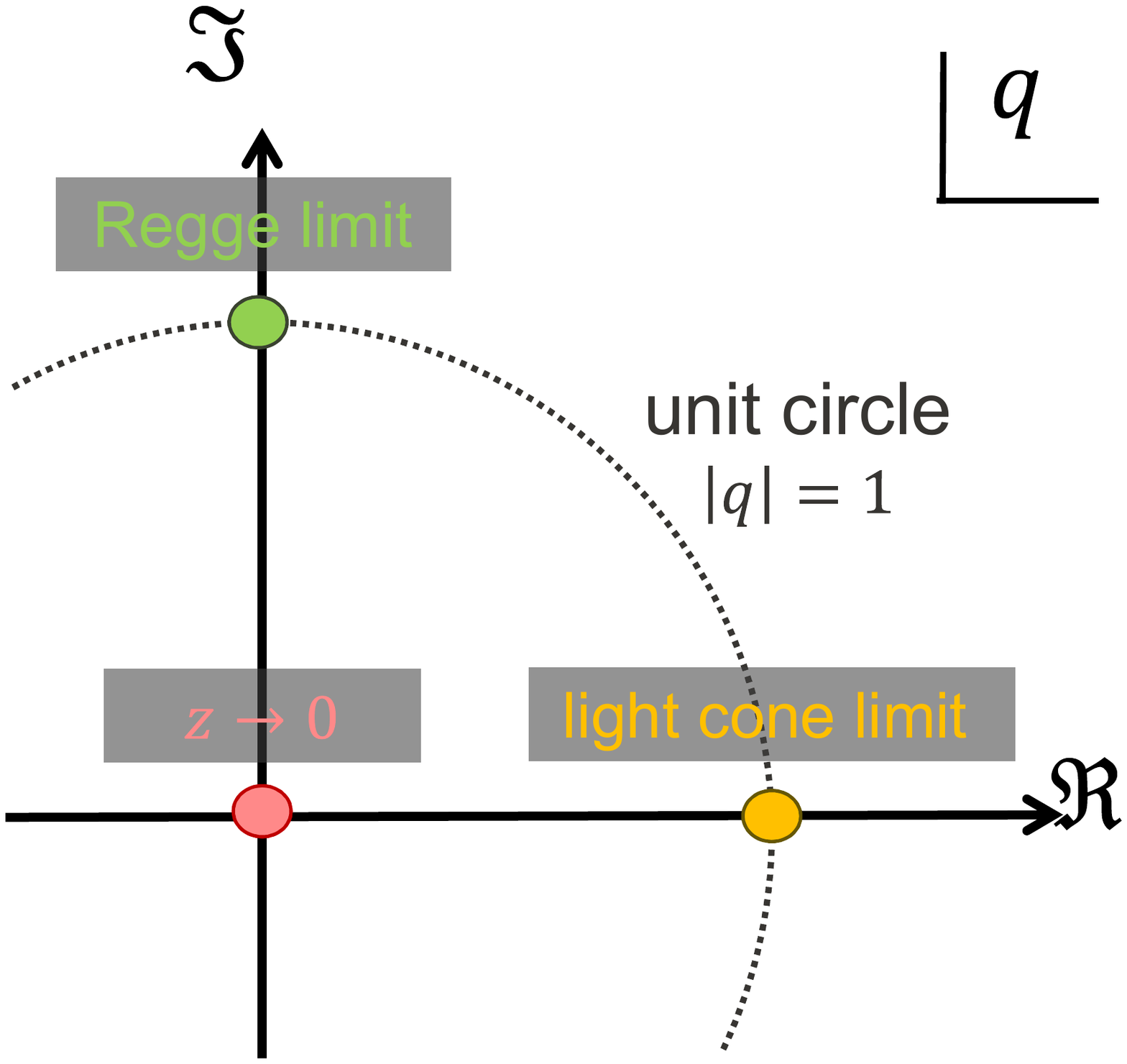}
 \end{center}
 \caption{The relation between cross ratio $z$ and elliptic nome $q$.}
 \label{fig:nome}
\end{figure}

The key point is that the Regge limit corresponds to the limit $q \to i, \bar{q} \to 0$ in terms of the elliptic nome (see Figure \ref{fig:nome}).
Considering that the series coefficients for $h_{A,B} >\fr{c-1}{32}$ are always non-negative and $z \to 1$ corresponds to $q \to 1$ and $z \to 0$ with $(1-z) \to \ex{-2\pi i} (1-z)$ corresponds to $q \to i$, we obtain the relationship between the light cone singularity and Regge singularity as follows:
\begin{equation}
\lim_{z \to 1} \abs{H^{21}_{34}(h_p|q)} \gsim  \lim_{\substack{z \to 0\\ \text{with}\\ (1-z)\to\ex{-2\pi i }(1-z) }} \abs{H^{21}_{34}(h_p|q)},
\end{equation}
where we only focus on the cases $(h_1,h_2,h_3,h_4)=(h_A,h_A,h_B,h_B)$ or $(h_A,h_B,h_B,h_A)$; however, we expect that this relationship can be generalized for any pairs with $h_{1,2,3,4}>\fr{c-1}{32}$.
Comparing (\ref{eq:preF}) with (\ref{eq:AABBs}), one can find that the main singularity in the light cone limit only comes from the prefactor  (\ref{eq:preF}),  and, therefore, the function $H^{21}_{34}(h_p|q)$ does not contribute to the singularity.
Therefore, we can conclude that
\begin{equation}
H^{21}_{34}(h_p|q)\ar{\substack{z \to 0\\ \text{with}\\ (1-z)\to\ex{-2\pi i }(1-z) }} O(\log z).
\end{equation}
This conclusion leads to the Regge singularity of a four-point function as follows:
\begin{equation}\label{eq:Regge}
G(z,\bar{z}) \ar{\substack{z \to 0\\ \text{with}\\ (1-z)\to\ex{-2\pi i }(1-z) }} z^{\fr{c-1}{24}-h_1-h_2} \bar{z}^{-h_1-h_2},
\ \ \ \ \ \text{if } h_{1,2,3,4}>\fr{c-1}{32}.
\end{equation}

One application of this result is the evaluation of the Renyi entropy after a local quench. As explained in Section \ref{subsec:2ndREE}, its growth can be analytically calculated using a four-point correlator with twist operators (\ref{eq:defREE}). In Section \ref{subsec:2ndREE}, we moved the framework from an orbifold CFT to a seed CFT using a conformal map to evaluate the growth; however, we can also calculate the growth using just (\ref{eq:defREE}) itself. In this calculation, we have to calculate the Regge limit of a four-point function, instead of the light cone limit. This result (\ref{eq:Regge}) leads to the conclusion that the Renyi entropy after a local quench shows universality if $n>n_* \text{ and } h_O>\fr{c-1}{32}$. This aspect is discussed in detail elsewhere \cite{Kusuki2018b}.
Note that the Regge limit also appears in the evaluation of OTOCs.

\section{Discussion}
The light cone structure of Virasoro conformal blocks allows us to access information about 2D CFTs.
For example, the large-spin spectrum can be derived from only the vacuum Virasoro conformal block through the light cone bootstrap equation.
Based on this background, an important task was to examine the light cone singularity of the Virasoro blocks in general.
In this paper, we reveal the light cone structure of general Virasoro blocks by investigating the fusion matrix (or the crossing kernel).
Interestingly, light cone singularity undergoes a phase transition as the external operator dimensions are varied. 
This fact leads to a richer structure of the spectrum at large spin; however, the physical interpretation of the transition is presently unclear.

At this stage, the most important future works are to understand the bulk interpretation of {\it universality of twist} and {\it transition at the BTZ threshold}. For this purpose, it might need to resolve the following problems:
\begin{quote}
i)	poor understanding of the relationship between Liouville CFT and $2+1$ dimensional gravity (see \cite{Krasnov2000, Krasnov2001, Jackson2015, Caputa2017b}),

ii)	dynamics of multiple deficit angles.
\end{quote}
The reason for considering (i) is because the transition point is characterised by the total Liouville momentum, instead of the total mass.

We expect that this transition would be related to thermalisation or black-hole formation. There are two reasons for this:
\begin{quote}
i)	The transition point is characterised by the {\it BTZ threshold}, similar to the thermalisation of the HHLL Virasoro blocks.

ii)	The transition can only be found in a CFT with $c>1$ and no conserved extra currents, which is expected to be chaotic. 
\end{quote}
In the first place, there is limited information as to why the BTZ threshold appears as the transition point. We intend to examine if the BTZ threshold is related to the creation of black hole.
It is interesting to explore the relationships between the transition, creation of a black hole, and thermalisation.

The light cone bootstrap equation suggests that there must be a universal binding energy even at large angular momentum. Moreover, this binding energy becomes larger if the total Liouville momentum increases beyond the BTZ threshold and the total twist is saturated by $\fr{c-1}{12}$ owing to the strong interaction above the BTZ threshold. It would be interesting to reproduce this binding energy at large spin from the calculation in AdS gravity.

The light cone singularity also reveals the entanglement structure. From our studies on entanglement in various setups, we found that the transition of the light cone singularity often caused discontinuousness of the derivative of the Renyi entropy in $n$. It is not possible to physically explain why the light cone limit destroys the assumption that the Renyi entropy is analytic in $n$. Apart from the transition in $n$, when considering the Renyi entropy after a local quench, we found that the Renyi entropy became large on increasing the conformal dimension of the operator used to create a quench, unless the dimension exceeded $\fr{c}{32}$.  Above the threshold, the Renyi entropy is saturated. We expect that it is related to the saturation of entanglement in some way. It would be interesting to explore this topic in future.

One interesting future work is to generalize the analytic bootstrap program \cite{Alday2017a, Simmons-Duffin2017} to two-dimensional CFTs. We believe that our result may contribute to this progress.  It would be interesting to explore this issue further.

\section*{Acknowledgments}

Especially, the author would like to express special thanks to Henry
Maxfield for a very helpful discussion. We are grateful to Tadashi Takayanagi for his fruitful discussions and comments. We also thank Masamichi Miyaji, Nilay Kundu, and Pawel Caputa for useful conversations about the subject of this study, and Jared Kaplan, Luis Fernando Alday, Nikita Nemkov and Taro Kimura for giving us useful comments.
YK is supported by the JSPS fellowship. 
YK is grateful to the conference ``Strings and Fields 2018'' in YITP, and the conference ``Recent Developments in Gauge Theory and String Theory'' in Keio U.

\appendix
\section{Light Cone Limit from Fusion Matrix} \label{app:FM}

\subsection{Leading Term in Light Cone Limit} \label{subapp:FM}

In the following, we introduce the notations usually found in Liouville CFTs.
\begin{equation}
c=1+6Q^2, \ \ \ \ \ Q=b+\fr{1}{b}, \ \ \ \ \ h_i=\a_i(Q-\a_i).
\end{equation}
Note that we can relate the parameter $\eta_i$ appearing in \cite{Kusuki2018a} to $\a_i$ as $\a_i=Q\eta_i$.

In this appendix, we show the asymptotic form of the conformal blocks in the limit $z\to 1$. The key point is that there are invertible fusion transformations between $s$ and $t$- channel conformal blocks \cite{Teschner2001a} as follows:
\footnote{A similar structure for a 1-pt function on a torus can be found in \cite{Teschner2003} (see also \cite{Hadasz2010b, Kraus2016, Nemkov2017}). It would be interesting to parallel our discussion for a 1-pt function on a torus.}
\begin{equation}\label{eq:fusiontrans}
\begin{aligned}
\ca{F}^{21}_{34}(h_{\a_s}|z)=\int_{\bb{S}} \dd \a_t {\bold F}_{\a_s, \a_t} 
   \left[
    \begin{array}{cc}
    \a_2   & \a_1  \\
     \a_3  &   \a_4\\
    \end{array}
  \right]
  \ca{F}^{23}_{14}(h_{\a_t}|1-z),
\end{aligned}
\end{equation}
where the contour $\bb{S}$ runs from $\fr{Q}{2}$ to $\fr{Q}{2}+ i\infty$. The kernel $ {\bold F}_{\a_s, \a_t} $ is called the {\it crossing matrix} or {\it fusion matrix}. The explicit form of the fusion matrix is given in \cite{Ponsot1999,Teschner2001a}as follows:
\begin{equation}\label{eq:crossing}
\begin{aligned}
{\bold F}_{\a_s, \a_t} 
   \left[
    \begin{array}{cc}
    \a_2   & \a_1  \\
     \a_3  &   \a_4\\
    \end{array}
  \right]
=\fr{N(\a_4,\a_3,\a_s)N(\a_s,\a_2,\a_1)}{N(\a_4,\a_t,\a_1)N(\a_t,\a_3,\a_2)}
   \left\{
    \begin{array}{cc|c}
    \a_1   & \a_2    &  \a_s  \\
     \a_3  & \a_4    &  \a_t   \\
    \end{array}
  \right\}_b,
\end{aligned}
\end{equation}
where the function $N(\a_3,\a_2,\a_1)$ is
\begin{equation}
N(\a_3,\a_2,\a_1)=\fr{\G_b(2\a_1)\G_b(2\a_2)\G_b(2Q-2\a_3)}{\G_b(2Q-\a_1-\a_2-\a_3)\G_b(Q-\a_1-\a_2+\a_3)\G_b(\a_1+\a_3-\a_2)\G_b(\a_2+\a_3-\a_1)},
\end{equation}
and 
$ \left\{
    \begin{array}{cc|c}
    \a_1   & \a_2    &  \a_s  \\
     \a_3  & \a_4    &  \a_t   \\
    \end{array}
  \right\}_b$
is the Racah--Wigner coefficient for the quantum group $U_q(sl(2,\bb{R}))$, which is given by
\footnote{Ponsot--Teschner have derived a more symmetric form of the Racah--Wigner coefficient \cite{Teschner2014} than the traditional expression found in \cite{Ponsot1999,Teschner2001a}. In this study, we used the new expression derived in \cite{Teschner2014}.}
\begin{equation}\label{eq:6j}
\begin{aligned}
&\left\{
    \begin{array}{cc|c}
    \a_1   & \a_2    &  \a_s  \\
     \a_3  & \bar{\a_4}    &  \a_t   \\
    \end{array}
  \right\}_b\\
&= \fr{S_b(\a_1+\a_4+\a_t-Q)S_b(\a_2+\a_3+\a_t-Q)S_b(\a_3-\a_2-\a_t+Q)S_b(\a_2-\a_3-\a_t+Q)}{S_b(\a_1+\a_2-\a_s)S_b(\a_3+\a_s-\a_4)S_b(\a_3+\a_4-\a_s)}\\
&\times \abs{S_b(2\a_t)}^2 \int^{2Q+i \infty}_{2Q-i \infty} \dd u 
\fr{S_b(u-\a_{12s})S_b(u-\a_{s34})S_b(u-\a_{23t})S_b(u-\a_{1t4})}{S_b(u-\a_{1234}+Q)S_b(u-\a_{st13}+Q)S_b(u-\a_{st24}+Q)S_b(u+Q)},
\end{aligned}
\end{equation}
where we have used the notations $\bar{\a}=Q-\a$, $\a_{ijk}=\a_i+\a_j+\a_k$ and $\a_{ijkl}=\a_i+\a_j+\a_k+\a_l$.
The functions $\G_b(x)$ and $S_b(x)$ are defined as
\begin{equation}
\G_b(x)= \fr{\G_2(x|b,b^{-1})}{\G_2\pa{\fr{Q}{2}|b,b^{-1}}}, \ \ \ \ \ S_b(x)=\fr{\G_b(x)}{\G_b(Q-x)},
\end{equation}
$\G_2(x|\w_1,\w_2)$ is the double gamma function,
\begin{equation}
\log \G_2(x|\w_1,\w_2)=\pa{\pd{t}\sum^{\infty}_{n_1,n_2=0} \pa{x+n_1 \w_1+n_2\w_2}^{-t}}_{t=0}.
\end{equation}
Note that the function $\G_b(x)$ is introduced such that $\G_b(x)=\G_{b^{-1}}(x)$ and satisfies the following relationship:
\begin{equation}
\G_b(x+b)=\fr{\s{2 \pi}b^{bx-\fr{1}{2}}}{\G(bx)}\G_b(x).
\end{equation}
By substituting the explicit form of the Racah--Wigner coefficients (\ref{eq:6j}) into (\ref{eq:crossing}), we can simplify the expression for the fusion matrix into
\begin{equation}\label{eq:crossing2}
\begin{aligned}
&{\bold F}_{\a_s, \a_t} 
   \left[
    \begin{array}{cc}
    \a_2   & \a_1  \\
     \a_3  &   \a_4\\
    \end{array}
  \right]\\
&=\fr{\G_b(Q+\a_2-\a_3-\a_t)\G_b(Q-\a_2+\a_3-\a_t)\G_b(2Q-\a_1-\a_4-\a_t)\G_b(\a_1+\a_4-\a_t)}{\G_b(2Q-\a_1-\a_2-\a_s)\G_b(\a_1+\a_2-\a_s)\G_b(Q+\a_3-\a_4-\a_s)\G_b(Q-\a_3+\a_4-\a_s)}\\
&\times \fr{\G_b(Q-\a_2-\a_3+\a_t)\G(-Q+\a_2+\a_3+\a_t)\G_b(\a_1-\a_4+\a_t)\G_b(-\a_1+\a_4+\a_t)}{\G_b(\a_1-\a_2+\a_s)\G_b(-\a_1+\a_2+\a_s)\G_b(Q-\a_3-\a_4+\a_s)\G_b(-Q+\a_3+\a_4+\a_s)}\\
&\times \abs{S_b(2\a_t)}^2 \fr{\G_b(2Q-2\a_s)\G_b(2\a_s)}{\G_b(2Q-2\a_t)\G_b(2\a_t)} \\
&\times \int^{2Q+i \infty}_{2Q-i \infty} \dd u 
\fr{S_b(u-\a_{12s})S_b(u-\a_{s34})S_b(u-\a_{23t})S_b(u-\a_{1t4})}{S_b(u-\a_{1234}+Q)S_b(u-\a_{st13}+Q)S_b(u-\a_{st24}+Q)S_b(u+Q)}.
\end{aligned}
\end{equation}

The conformal blocks have the following simple asymptotic form:
\begin{equation}
\ca{F}^{21}_{34}(h_{\a_s}|z) \ar{z\to 0} z^{h_s-h_1-h_2}(1+O(z)).
\end{equation}
Naively substituting this asymptotics into the fusion transformation (\ref{eq:fusiontrans}) leads to
\begin{equation}
\ca{F}^{21}_{34}(h_{\a_s}|z) \ar{z\to 1} (1-z)^{\fr{c-1}{24}-h_2-h_3}.
\end{equation}
Interestingly, this reproduces a part of the asymptotic formulas (\ref{eq:AABBs}) and (\ref{eq:ABBAs}). However, we cannot straightforwardly understand how the transition seen in (\ref{eq:AABBs}) and (\ref{eq:ABBAs}) can be reproduced using the transformation (\ref{eq:fusiontrans}).

\begin{figure}[t]
 \begin{center}
  \includegraphics[width=15.0cm,clip]{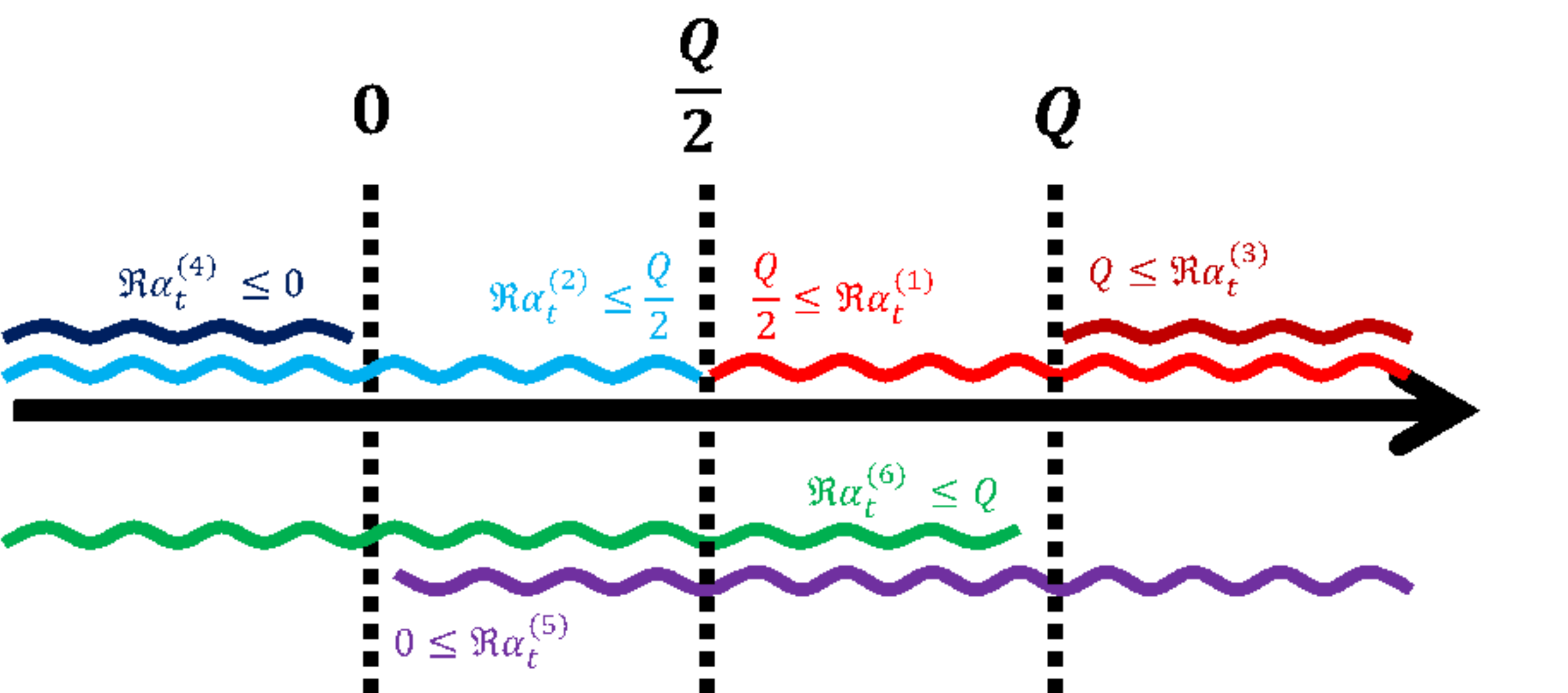}
 \end{center}
 \caption{The domain of $\Re \a_t^{(i)}$. From this figure, we can find that only $\a_t^{(5)}$ and $\a_t^{(6)}$ can cross the contour $\bb{S}$ of the integral over $\a_t$.}
 \label{fig:alpha}
\end{figure}

\begin{figure}[t]
 \begin{center}
  \includegraphics[width=10.0cm,clip]{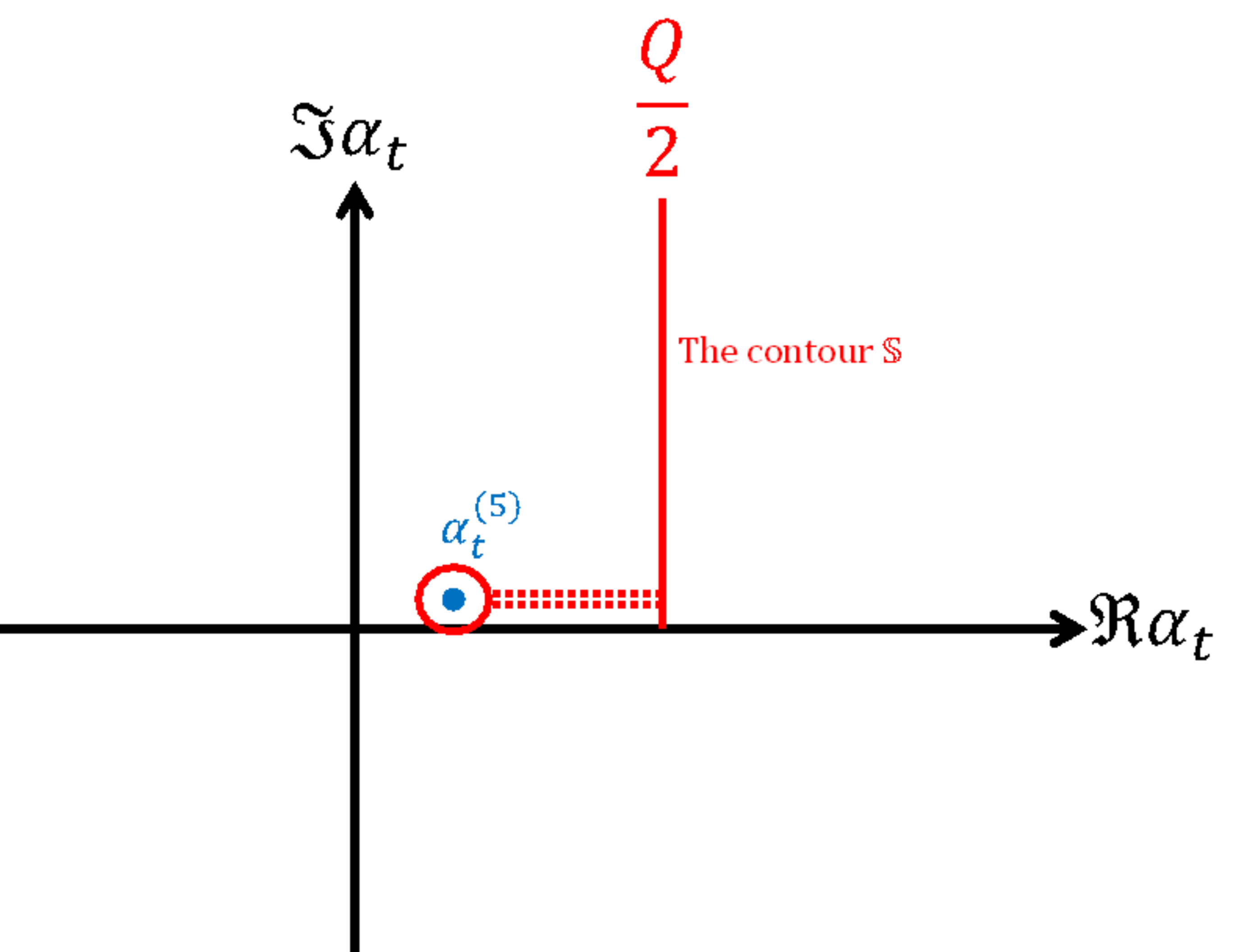}
 \end{center}
 \caption{The contour $\bb{S}$ deformed by the pole $\a_t^{(5)}$.}
 \label{fig:contour}
\end{figure}

The key to understanding how the transition is derived from the expression of the fusion matrix is that the function $\G_b(x)$ has poles at $x=-\pa{mb+\fr{n}{b}}$ for $n,m \in \bb{Z}_{\geq0}$, and therefore, the fusion matrix also has poles. The following shows the poles in the first and second lines of the expression (\ref{eq:crossing2}):

[{\it Poles of the Numerator}]
\begin{equation}\label{eq:poles}
\begin{aligned}
\a_t^{(1)}&=\pm(\a_2-\a_3)+Q+Q_{m,n}\\
\a_t^{(2)}&=\pm(\a_1-\a_4)-Q_{m,n}\\
\a_t^{(3)}&=-(\a_1+\a_4)+2Q+Q_{m,n}\\
\a_t^{(4)}&=(\a_2+\a_3)-Q-Q_{m,n}\\
\a_t^{(5)}&=\a_1+\a_4+Q_{m,n}\\
\a_t^{(6)}&=-(\a_2+\a_3)+Q-Q_{m,n}\\
\end{aligned}
\end{equation}
where we define $Q_{m,n}=mb+\fr{n}{b}$ for $n,m \in \bb{Z}_{\geq0}$. The real values of $\a_i$ always satisfy $0 \leq \Re \a_i \leq \fr{Q}{2}$ for the operators in unitary CFTs, and the value $Q_{m,n}$ is always positive by definition. Therefore, each value of $\Re \a_t^{(i)}$ exists only in a particular domain as shown in Figure \ref{fig:alpha}. One can find that only $\a_t^{(5)}$ and $\a_t^{(6)}$ can cross the contour $\bb{S}$ of the integral over $\a_t$ when $\a_1+\a_4<\fr{Q}{2}$. As a result of this crossing,  the contour $\bb{S}$ is deformed as shown in Figure \ref{fig:contour}. 
\footnote{This deformation of the contour in a particular case was described in \cite{Chang2016} and we are very much grateful to Henry Maxfield for pointing out this.}

This deformation leads to another possibility that the conformal block has the asymptotics as follows:
\begin{equation}
\begin{aligned}
\ca{F}^{21}_{34}(h_{\a_s}|z)& \ar{z\to 1} \left\{
    \begin{array}{ll}
     (1-z)^{h_1+h_4-h_2-h_3-2\a_1 \a_4}  ,& \text{if } \a_1+\a_4<\fr{Q}{2}   ,\\
     (1-z)^{-2\a_2\a_3}  ,& \text{if }  \a_2+\a_3<\fr{Q}{2} .\\
    \end{array}
  \right.\\
\end{aligned}
\end{equation}
If we set three parameters as
\begin{equation}
\kappa_1=h_1+h_4-h_2-h_3-2\a_1 \a_4, \ \ \ \ \ 
\kappa_2=-2\a_2\a_3, \ \ \ \ \ 
\kappa_3=\pa{\fr{c-1}{24}-h_1-h_4},
\end{equation}
then we can show the following inequalities:
\begin{equation}
\begin{aligned}
\kappa_3-\kappa_1&=\pa{\a_1+\a_4-\fr{Q}{2}}^2>0,\\
\kappa_3-\kappa_2&=\pa{\a_2+\a_3-\fr{Q}{2}}^2>0,\\
\kappa_1-\kappa_2&=(\a_1+\a_4-\a_2-\a_3)(1-\a_1-\a_2-\a_3-\a_4),
\end{aligned}
\ \ \ \ \ 
\begin{aligned}
&\text{if }\a_1+\a_4<\fr{Q}{2},\\
&\text{if }\a_2+\a_3<\fr{Q}{2},\\
&\text{if }\a_1+\a_4<\fr{Q}{2},  \a_2+\a_3<\fr{Q}{2}.\\
\end{aligned}
\end{equation}
Therefore, the leading contribution of the ABBA blocks in the limit $z \to 1$ is given by
\begin{equation}\label{eq:FMresult1}
\begin{aligned}
\ca{F}^{BA}_{BA}(h_{\a_s}|z)& \ar{z\to 1} \left\{
    \begin{array}{ll}
     (1-z)^{4h_A-2h_B-2Q\a_A}  ,& \text{if } \a_A<\fr{Q}{4}\ \text{and } \ \a_A<\a_B   ,\\
     (1-z)^{2h_B-2Q\a_B}  ,& \text{if } \a_B<\fr{Q}{4}\ \text{and } \ \a_B<\a_A   ,\\
     (1-z)^{\fr{c-1}{24}-2h_B}  ,& \text{otherwise } .\\
    \end{array}
  \right.\\
\end{aligned}
\end{equation}
Further, the asymptotics of the AABB blocks is given by
\begin{equation}\label{eq:FMresult2}
\begin{aligned}
\ca{F}^{AA}_{BB}(h_{\a_s}|z)& \ar{z\to 1} \left\{
    \begin{array}{ll}
     (1-z)^{-2\a_A\a_B}  ,& \text{if } \a_A+\a_B<\fr{Q}{2}\   ,\\
     (1-z)^{\fr{c-1}{24}-h_A-h_B}  ,& \text{otherwise }  .\\
    \end{array}
  \right.\\
\end{aligned}
\end{equation}
Interestingly, these results exactly match our previous results obtained using the monodromy method in \cite{Kusuki2018a} and numerical results in \cite{Kusuki2018,Kusuki2018b}. We intend to emphasize that the above calculation based on the fusion matrix does not rely on the large $c$ limit. Therefore, we expect that the asymptotic form derived from the monodromy method with $c \to \infty$ in \cite{Kusuki2018a} can be generalized to any unitary CFTs with $c>1$ by replacing the factor $c$ by $c-1$ .

One might doubt this result because it contradicts with the singularities appearing in the minimal models. In the first place, when one of the external operators corresponds to a degenerate operator, the decomposition of the $s$-channel in terms of the $t$-channel is not a continuum, but is discrete. That is, in such a case, the crossing kernel needs to be written as a linear combination of delta functions. In fact, if the external operator is degenerate, $S_b$ or $\G_b$ in the denominator of the expression (\ref{eq:crossing2}) diverges, and therefore, the crossing kernel vanishes for general $\a_t$; however, these divergences can be cancelled only if $\a_t$ takes particular values. As a result, the $s$-channel is decomposed as a discrete sum of the $t$-channels, instead of an integral over the continuum spectrum. This concept is explained in greater detail in \cite{Hadasz2005,EsterlisFitzpatrickRamirez2016}. We have to mention that this never happens for unitary CFTs with $c>1$ because if the central charge is larger than one, then the conformal dimensions of degenerate operators are negative.

\subsection{Sub-leading Terms in Light Cone Limit} \label{subapp:sub}

In Section \ref{subsec:spectrum}, we tried to derive the spectrum of twist. For this, we need the sub-leading contributions of the conformal blocks in the light cone limit. In this appendix, we first give the sub-leading terms in the semiclassical limit using the fusion matrix and then generalize it.

\begin{figure}[t]
 \begin{center}
  \includegraphics[width=9.0cm,clip]{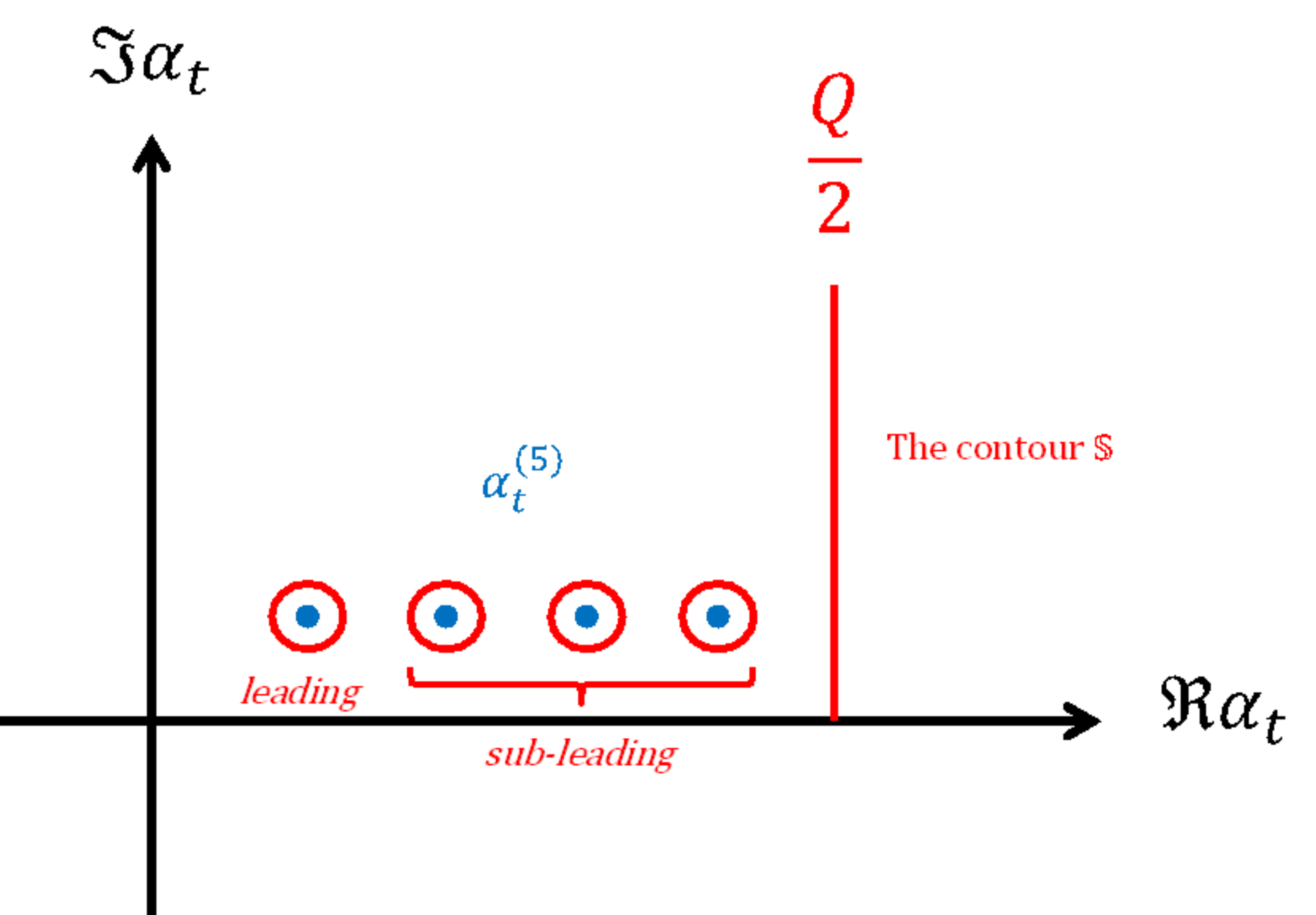}
 \end{center}
 \caption{This figure shows from where the sub-leading contributions to the blocks are obtained.}
 \label{fig:contour2}
\end{figure}

We expect that the sub-leading terms come from the poles (\ref{eq:poles}) with non-zero $n,m$.
As an example, let us consider the contributions from the poles $\a_t^{(5)}$ with non-zero $n,m$ in the semiclassical limit.
If $\a_1+\a_4+Q_{m,n}<\fr{Q}{2}$ is satisfied, the corresponding term labelled $m,n$ also contributes to the conformal blocks as sub-leading terms. Its singularity is given by
\begin{equation}
(1-z)^{h_1+h_4-h_2-h_3-2\a_1\a_4+\m_{m,n}},
\end{equation}
where $\m_{m,n}$ is a positive constant defined by
\begin{equation}
\m_{m,n}=\pa{mb+\fr{n}{b}} \pa{(1-\w_{m,n}+m)b+\fr{1-\w_{m,n}+n}{b}},
\end{equation}
and $\w_{m,n} \equiv \fr{2}{Q}(\a_1+\a_4+Q_{m,n})<1$. If taking the limit $b \to 0$ with $h_1, h_4/c$ fixed, the sub-leading terms are given by $\m_{m,0}$ as $\m_{m,n>0} \gg \m_{m,0}$, and $\w$ can be approximated as $\w_{m,0} \simeq 1-\d$ with $\d=\s{1-\fr{24}{c}h_4}$.  Thus, we can approximate $\m_{m,0}$ as
\begin{equation}
\m_{m,0} \ar{b\to0} m\d.
\end{equation}
This leads to the sub-leading contributions to the conformal blocks as follows:
\begin{equation}\label{eq:subHHLL}
\ca{F}^{LL}_{HH}(h_p|z)\ar{z\to1}  \sum_{m \in \bb{Z}_{\geq0} } \ca{P}_m (1-z)^{\d(h_L+m)-h_L},
\end{equation}
where $\ca{P}_m$ are some constants. In fact, the HHLL Virasoto blocks have the same form.
\begin{equation}\label{eq:fullHHLL}
\begin{aligned}
\ca{F}^{LL}_{HH}(h_p|z)&= (1-z)^{h_L(\d-1)}\pa{\fr{1-(1-z)^\d}{\d}}^{h_p-2h_L} {}_2F_1(h_p,h_p,2h_p|1-(1-z)^\d)\\
&\ar{z\to1}  \sum_{m \in \bb{Z}_{\geq0} } \ca{P}_m (1-z)^{\d(h_L+m)-h_L},
\end{aligned}
\end{equation}
which can be easily checked using the series expansion of $\pa{1-x}^{-2h_L}$.
Therefore, we can conclude that the sub-leading contributions come from the poles with non-zero $m,n$, as shown in Figure \ref{fig:contour2}.
Note that now we take the limit $b \to 0$; therefore, an infinite number of poles $\a_1+\a_4+Q_{m,0}$ for any $m \in \bb{Z}_{\geq0}$ satisfy the inequality 
\begin{equation}
\a_1+\a_4+Q_{m,0}<\fr{Q}{2},
\end{equation}
when the inequality  $\a_1+\a_4<\fr{Q}{2}$ is satisfied. Thus, an infinite number of poles contribute to the conformal blocks as shown in (\ref{eq:fullHHLL}) in the semiclassical limit.
 The condition $\fr{h_1}{c} \ll1$ can immediately be relaxed to non-perturbative $\fr{h_1}{c}$ as
\begin{equation}
\begin{aligned}
\ca{F}^{AA}_{BB}(h_p|z)&\ar{z\to1} \left\{
    \begin{array}{ll}
        \sum_{m \in \bb{Z}_{\geq0} } \ca{P}_m (1-z)^{-2\a_A \a_B +m\pa{1-\fr{2}{Q}(\a_A+\a_B)}},& \text{if } \a_A+\a_B<\fr{Q}{2},\\
       (1-z)^{\fr{c-1}{24}-h_A-h_B}, & \text{otherwise}   .\\
    \end{array}
  \right.\\
\end{aligned}
\end{equation}

For general unitary CFTs with finite $c>1$, there are only finite poles satisfying the inequality $\a_1+\a_4+Q_{m,n}<\fr{Q}{2}$.
Therefore, the conformal blocks can be expressed using a finite sum as
\begin{equation}\label{eq:fullAABB}
\ca{F}^{AA}_{BB}(h_p|z)\ar{z\to1}  \sum_{\substack{m,n \in \bb{Z}_{\geq0}\\ \text{where} \\ \a_A+\a_B+Q_{m,n}<\fr{Q}{2} }}
 \ca{P}_{m,n} (1-z)^{-2\a_A \a_B+\m_{m,n}}, \ \ \ \ \  \text{if } \a_A+\a_B<\fr{Q}{2},
\end{equation}
and 
\begin{equation}
\ca{F}^{AA}_{BB}(h_p|z)\ar{z\to1}  (1-z)^{\fr{c-1}{24}-h_A-h_B}, \ \ \ \ \  \text{otherwise },
\end{equation}
where $\ca{P}_{m,n}$ are some constants. Note that the above process can straightforwardly be applied to the ABBA blocks.
That is, the result can be obtained by just adding $\m_{m,n}$ to the power of its singularity.

Many studies on the fusion matrix approach remain to be carried out. The most important one is identifying the logarithmic corrections to the light cone limit of the conformal blocks. This correction is related to the other anomalous dimension different from (\ref{eq:Ebound}), which also appears in higher dimensions.

\section{Light Cone Modular Bootstrap} \label{app:modular}

\subsection{Light Cone Modular Bootstrap}

The modular invariance imposes the following condition on the torus partition function:
\begin{equation}\label{eq:modular}
Z(\tau,\bar{\tau})=Z\pa{-\fr{1}{\tau},-\fr{1}{\bar{\tau}}},
\end{equation}
where the torus partition function has the Virasoro character decomposition in CFTs,
\begin{equation}
Z(\tau,\bar{\tau})=\sum_{i,j} D_{i,j} \chi_{h_i}(\tau)\bar{\chi}_{\bar{h}_j}(\bar{\tau})
=\int \dd h \dd \bar{h} \ \ \rho(h,\bar{h}) \chi_h(\tau)\bar{\chi}_{\bar{h}}(\bar{\tau}).
\end{equation}
Here, function $D_{i,j}$ denotes the degeneracy of primary operators of weight $(h_i,\bar{h}_j)$, and function $\rho (h,\bar{h})$ is defined as $\rho (h,\bar{h})=\sum_{i,j} D_{i,j} \delta(h-h_i) \delta(\bar{h}-\bar{h}_j)$.
If we limit ourselves to CFTs with $c>1$, the Virasoro character has the following simple form:
\begin{equation}
\chi_0(\tau)=\fr{q^{-\fr{c-1}{24}}}{\eta(\tau)}(1-q), \ \ \ \  \chi_h(\tau)=\fr{q^{h-\fr{c-1}{24}}}{\eta(\tau)},
\end{equation}
where $q=\ex{2\pi i \tau}$ and $\eta(\tau)$ represents the Dedekind eta function.

Here, we focus on CFTs with $c>1$ and without extra currents.
The partition function of such CFTs in the limit $q \to 0$ (or equivalently $\tau \to i \infty$) can be approximated by the vacuum contribution:
\begin{equation}
Z(\tau,\bar{\tau}) \ar{q \to 0} \chi_0(\tau) \chi_0(\bar{\tau}).
\end{equation}
If we define the {\it large spin spectrum} as
\begin{equation}
\rho (\infty, \bar{h}) \equiv \lim_{\tau \to i\infty}  \int \dd h \ \ \rho(h,\bar{h}) \fr{\chi_h \pa{-\fr{1}{\tau}}}{\chi_0 (\tau)},
\end{equation}
then we can factor out the $\bar{\tau}$ and $\bar{z}$ dependence in the modular bootstrap equation as
\begin{equation}\label{eq:chiral modular}
\chi_0 (\bar{\tau})=\int \dd \bar{h} \ \ \rho(\infty,\bar{h}) \bar{\chi}_{\bar{h}}\pa{-\fr{1}{\bar{\tau}}}.
\end{equation} 
The S-transformation of the vacuum character can be expressed as \cite{Zamolodchikov2001}
\begin{equation}\label{eq:chiral modular2}
\chi_0 (\bar{\tau})=\int_{\fr{c-1}{24}}^{\infty} \dd \bar{h} \ \ S(\bar{h},0) \bar{\chi}_{\bar{h}}\pa{-\fr{1}{\bar{\tau}}},
\end{equation} 
where the S-matrix is 
\begin{equation}
\begin{aligned}
S(\bar{h},0)=2\s{2}\pa{\bar{h}-\fr{c-1}{24}}^{-\fr{1}{2}} \sinh \pa{2 \pi b \s{\bar{h}-\fr{c-1}{24}}} \\
\times \sinh \pa{2 \pi b^{-1} \s{\bar{h}-\fr{c-1}{24}}}.
\end{aligned}
\end{equation}
Upon comparing (\ref{eq:chiral modular}) and (\ref{eq:chiral modular2}), we find that the large spin spectrum is given by
\begin{equation}\label{eq:large spin spectrum}
\rho(\infty,\bar{h})=S(\bar{h},0).
\end{equation}
In particular, the asymptotic behavior of this spectrum reproduces the Cardy formula \cite{Cardy1986a},
\begin{equation}
\rho(\infty,\bar{h}) \ar{\bar{h} \to \infty} \ex{2\pi \s{\fr{c-1}{6} \pa{\bar{h}-\fr{c-1}{24}}}}
\end{equation}
 The result (\ref{eq:large spin spectrum}) suggests that such CFTs have a universal spectrum at large spin.

On the other hand, if we assume that there are only primary operators with bounded spin in a CFT, then we can show that the lowest-lying operator in such a CFT must have the dimension $h_{min}=\fr{c-1}{24}$, in that this CFT is non-compact (see \cite{Collier2016} for more details). 

The cosmic censorship conjecture requires the condition $\abs{J} \leq M$, which is translated in terms of conformal dimensions as
\begin{equation}
\min \{h,\bar{h} \} \geq \fr{c}{24}.
\end{equation}
In particular, a state with $\min \{h,\bar{h} \} = \fr{c}{24}$ corresponds to an extremal BTZ black hole (e.g., Strominger--Vafa black hole \cite{Strominger1996}).
In this sense, the large spin spectrum  (\ref{eq:large spin spectrum}) corresponds to BTZ black holes.

\subsection{Relation between Conformal and Modular Bootstrap}\label{subapp:relation}
The counterpart of the light cone modular bootstrap (\ref{eq:chiral modular}) can be given by

\begin{equation}\label{eq:bootstrapAABB}
\overline{\ca{F}^{AA}_{BB}}(0|\bar{z})=\int  \dd \bar{h}  \ \ \rho_{AB} (\infty,\bar{h}) \overline{\ca{F}^{AB}_{AB}}(\bar{h}|1-\bar{z}),
\end{equation}
where we define the {\it large spin spectral density},
\begin{equation}
\rho_{AB} (\infty,\bar{h}) \equiv \lim_{z \to 0} \int \dd h  \ \ \rho_{AB} (h,\bar{h}) \fr{ \ca{F}^{AB}_{AB}(h|1-z)}{\ca{F}^{AA}_{BB}(0|z)}.
\end{equation}
The counterpart of (\ref{eq:chiral modular2}) is given by the fusion matrix,
\begin{equation}\label{eq:bootstrapAABB2}
\begin{aligned}
&\ca{F}^{AA}_{BB}(0|z)\\
&=
-2\pi i  \sum_{\substack{\a_{n,m}<\fr{Q}{2} \\ n,m \in \bb{Z}_{\geq0}}}\ \text{Res}\pa{   
  {\bold F}_{0, \a_t} 
   \left[
    \begin{array}{cc}
    \a_A   & \a_A  \\
     \a_B  &   \a_B\\
    \end{array}
  \right]
  \ca{F}^{AB}_{AB}(h_{\a_t}|1-z);\a_{n,m}}\\
&+
\int_{\fr{Q}{2}+0}^{\fr{Q}{2}+i \infty} \dd \a_t {\bold F}_{0, \a_t} 
   \left[
    \begin{array}{cc}
    \a_A   & \a_A  \\
     \a_B  &   \a_B\\
    \end{array}
  \right]
  \ca{F}^{AB}_{AB}(h_{\a_t}|1-z),
\end{aligned}
\end{equation}
where $\a_{n,m}\equiv\a_A+\a_B+mb+nb^{-1}$.
Since the asymptotic behavior of the Virasoro blocks is given by
\begin{equation}
\ca{F}^{ji}_{kl}(h_p|z) \ar{z \to 0} z^{h_p-h_i-h_j}
\end{equation}
if we take the limit $\bar{z} \to 1$ in (\ref{eq:bootstrapAABB}), then we obtain the twist spectrum and the structure constant at large spin from the fusion matrix  (\ref{eq:bootstrapAABB2}), which is the so-called {\it light-cone bootstrap}.
\footnote{ A very similar approach can be found in \cite{Francesco2012} (see (2.115)$\sim$(2.117)), which derives a general relation between the OPE coefficients and the fusion matrices.
}
In fact, upon comparing (\ref{eq:bootstrapAABB}) and (\ref{eq:bootstrapAABB2}), we find that the twist spectrum in the OPE between $O_A$ and $O_B$ at large spin can be expressed as
\begin{equation}\label{eq:LCresult}
\begin{aligned}
 \bar{h}&=\left\{
    \begin{array}{ll}
   \{ \bar{\a}_{n,m}(Q-\bar{\a}_{n,m}) \ \ \ \text{for $n,m \in \bb{Z}_{\geq 0}$ s.t. $\bar{\a}_{n,m}<\fr{Q}{2}$} \} \cup \kagi{\fr{c-1}{24},\infty}, \\
	\hspace{5cm} (\text{if }  \bar{\a}_A+\bar{\a}_B<\fr{Q}{2}) ,\\ \\
    \kagi{  \fr{c-1}{24},\infty}, \hspace{3.7cm} ( \text{if } \bar{\a}_A+\bar{\a}_B\geq\fr{Q}{2})  ,\\
    \end{array}
  \right.\\
\end{aligned}
\end{equation}
where $\bar{\a}_{n,m}\equiv\bar{\a}_A+\bar{\a}_B+mb+nb^{-1}$.

This spectrum has a simple interpretation. If we consider the analytic continuation of the Liouville OPE to $\a\notin ]0,Q[$, we can obtain the following fusion rules \cite{Ribault2014},
\begin{equation}\label{eq:fusion rule}
\begin{aligned}
\ca{V}_{\a_A} \times \ca{V}_{\a_B} =
 \sum_{\substack{\a_{n,m}<\fr{Q}{2} \\ n,m \in \bb{Z}_{\geq0}}} \ca{V}_{\a_{n,m}}
+ \int^{\fr{Q}{2}+i \infty}_{\fr{Q}{2}+0} \dd \a \ \ca{V}_{\a},
\end{aligned}
\end{equation}
where $\ca{V}_\a$ denotes a primary operator characterized  by a conformal dimension $h=\a(Q-\a)$.
Upon inspecting this fusion rule, we can straightforwardly observe that the twist spectrum (\ref{eq:LCresult}) exactly matches the spectrum originating from the fusion between primary operators in the (extended) Liouville CFT (\ref{eq:fusion rule}). Therefore, we can conclude that 
the twist spectrum at large spin approaches that of the {\it (extended) Liouville CFT}. This is the Virasoro counterpart of the statement that the twist spectrum at large spin in higher-dimensional CFTs ($d\geq3$) approaches the spectrum of double-trace states in a {\it generalized free theory}. 

The Liouville CFT sometimes appears when we study chiral solutions to pure 3D gravity, which can describe spinning two-particle states in AdS${}_3$  \cite{Hulik2017, Hulik2018}. This spinning two-particle solution leads to the energy $\bar{h}=\bar{\a}_{0,0}(Q-\bar{\a}_{0,0})$.
In this sense, we can reproduce the large spin twist spectrum from the bulk. This solution is described by the Liouville gravity, therefore,  it is naturally expected that at the quantum level, the spectrum of this two-particle state can be described by the fusion rules of the extended Liouville CFT (\ref{eq:fusion rule}), and therefore, we can conclude that the twist spectrum of this two-particle state is given as $\bar{h}=\bar{\a}_{n,m}(Q-\bar{\a}_{n,m}), \ \ \ n,m\in \bb{Z}_{\geq 0} \ \ \text{s.t.} \ \ \bar{h}<\fr{c-1}{24}$.

\section{Zamolodchikov Recursion Relation} \label{app:recursion}
The Zamolodchikov recursion relation \cite{Zamolodchikov1987,Zamolodchikov1984} is one of the tools used to calculate the conformal blocks numerically and has been recently receiving much attention \cite{Chen2017,Ruggiero2018} because it effectively encompasses the conformal blocks beyond the known regimes or limits. We provide a brief explanation herein.
\footnote{The Zamolodchikov recursion relation is also used in the conformal bootstrap \cite{EsterlisFitzpatrickRamirez2016, BaeLeeLee2016, LinShaoSimmons-DuffinWangYin2017,CollierKravchukLinYin2017}. An earlier study \cite{Perlmutter2015} presents a good review of this concept and discusses the connections between various recursion relations.
 A generalization of the recursion relation to more general Riemann surfaces is given in \cite{Cho2017a}.
}

For simplicity, we will decompose the conformal blocks into two parts as
\begin{equation}
\ca{F}^{21}_{34}(h_p|z)=\Lambda^{21}_{34}(h_p|q)H^{21}_{34}(h_p|q),\ \ \ \ \ \ q(z)=\ex{-\pi \fr{K(1-z)}{K(z)}},
\end{equation}
where $K(x)$ is the elliptic integral of the first kind and the function $\Lambda^{21}_{34}(h_p|q)$ is a universal prefactor given by
\begin{equation}\label{eq:pre}
 \Lambda^{21}_{34}(h_p|q)=(16q)^{h_p-\frac{c-1}{24}}z^{\frac{c-1}{24}-h_1-h_2}(1-z)^{\frac{c-1}{24}-h_2-h_3}
(\theta_3(q))^{\frac{c-1}{2}-4(h_1+h_2+h_3+h_4)}.
\end{equation}
The function $H^{21}_{34}(h_p|q)$ can be calculated recursively using the following relation:
\begin{equation}
H^{21}_{34}(h_p|q)=1+\sum^\infty_{m=1,n=1}\frac{q^{mn}R_{m,n}}{h_p-h_{m,n}}H^{21}_{34}(h_{m,n}+mn|q),
\end{equation}
where $R_{m,n}$ is a constant in $q$, which is defined by
\begin{equation}\label{eq:Rmn}
R_{m,n}=2\fr{
\substack{m-1\\ \displaystyle{\prod} \\p=-m+1\\p+m=1 (\text{mod } 2) \  } \ 
\substack{n-1\\ \displaystyle{\prod} \\q=-n+1\\q+n=1 (\text{mod } 2) }
\pa{\la_2+\la_1-\la_{p,q}}\pa{\la_2-\la_1-\la_{p,q}}\pa{\la_3+\la_4-\la_{p,q}}\pa{\la_3-\la_4-\la_{p,q}}}
{\substack{
\substack{m \\ \displaystyle{\prod} \\k=-m+1 } \ \ 
\substack{n \\ \displaystyle{\prod} \\l=-n+1 }\\
(k,l)\neq(0,0), (m,n)
}
 \la_{k,l}}.
\end{equation}
In the above expressions, we used the notations
\begin{equation}
\begin{aligned}
&c=1+6\pa{b+\fr{1}{b}}^2,  \hspace{16ex}   h_i=\fr{c-1}{24}-\l_i^2,\\
&h_{m,n}=\fr{1}{4}\pa{b+\fr{1}{b}}^2-\lambda_{m,n}^2,  \hspace{10ex}  \lambda_{m,n}=\fr{1}{2} \pa{\fr{m}{b}+nb}.
\end{aligned}
\end{equation}
Unfortunately, this recursion process is too complicated for calculating the Virasoro blocks analytically; however, from the viewpoint of numerical computations, this recursion relation is much more useful than the BPZ method \cite{Belavin1984}. 
One of the recent results involving the recursion relation is presented in our previous papers \cite{Kusuki2018b, Kusuki2018}. It reveals the general solutions to this recursion relation via numerical computations. For simplicity, we re-express the function $H^{21}_{34}(h_p|q)$ as
\be
H^{21}_{34}(h_p|q)=1+\sum_{k=1}^\infty c_k(h_p) q^{k},
\ee
and the corresponding recursion relation as
\begin{equation}\label{eq:ck}
	c_k(h_p) = \sum_{i=1}^k \sum_{\substack{m=1, n=1\\mn=i}} \frac{R_{m,n}}{h_p - h_{m,n}} c_{k-i}(h_{m,n}+mn).
\end{equation}
For this series expansion form, our previous numerical computations suggest that the solution $c_n$ for large $n$ takes the simple {\it Cardy-like} form of
\begin{equation}\label{eq:cn}
c_n \sim\xi_n  n^{\a} \ex{A\s{n}} \ \ \ \ \ \ \ \ \text{for large $n\gg c$},
\end{equation}
where 
\begin{equation}\label{eq:xin}
\begin{aligned}
\xi_n&=\left\{
    \begin{array}{ll}
      \d_{n,\text{even}} \times \sgn\BR{\pa{h_A-\fr{c-1}{32}}\pa{h_B-\fr{c-1}{32}}}^{\fr{n}{2}} ,& \text{for }  \ca{F}^{AA}_{BB}(h_p|z)   ,\\
      1 ,& \text{for } \ca{F}^{BA}_{BA}(h_p|z)   .\\
    \end{array}
  \right.\\
\end{aligned}
\end{equation}
The parameters $A$ and $\a$ are non-trivial, depending on the external conformal dimensions and the central charge.
\begin{quote}
{\large \it AABB blocks:}
\begin{equation}\label{eq:AABBA}
\begin{aligned}
A&=\left\{
    \begin{array}{ll}
    2\pi\s{\fr{c-1}{24}-h_A-h_B+2\a_A\a_B}   ,& \text{if } h_A.h_B>\fr{c-1}{32}  ,\\
    \pi\s{\fr{c-1}{24}-2h_A}   ,& \text{if } h_B>\fr{c-1}{32}>h_A  ,\\
   0  ,& \text{if } \fr{c-1}{32}>h_A.h_B  ,\\
    \end{array}
  \right.\\
\end{aligned}
\end{equation}

\begin{equation}\label{eq:AABBal}
\begin{aligned}
\a&=\left\{
    \begin{array}{ll}
    2(h_A+h_B)-\fr{c+5}{8}   ,& \text{if } h_A.h_B>\fr{c-1}{32}  ,\\
    4(h_A+h_B)-\fr{c+9}{4}  ,& \text{otherwise }   .\\
    \end{array}
  \right.\\
\end{aligned}
\end{equation}

{\large \it ABBA blocks:} \\
(For simplicity, we assume $h_B>h_A$, but it does not matter.)
\begin{equation}\label{eq:ABBAA}
\begin{aligned}
A&=\left\{
    \begin{array}{ll}
    2\pi\s{\fr{c-1}{24}-4h_A+2Q\a_A}   ,& \text{if } h_A<\fr{c-1}{32}  ,\\
   0   ,& \text{if } h_A>\fr{c-1}{32}  ,\\
    \end{array}
  \right.\\
\end{aligned}
\end{equation}

\begin{equation}\label{eq:ABBAal}
\begin{aligned}
\a&=\left\{
    \begin{array}{ll}
    2(h_A+h_B)-\fr{c+5}{8}   ,& \text{if } h_A<\fr{c-1}{32} ,\\
    4(h_A+h_B)-\fr{c+9}{4} ,& \text{if } h_A>\fr{c-1}{32}   .\\
    \end{array}
  \right.\\
\end{aligned}
\end{equation}
\end{quote}
These expressions are numerically conjectured in \cite{Kusuki2018b, Kusuki2018} and partly proven in a particular case in \cite{Kusuki2018a}.

In the limit $z \to 1$, the summation can be approximated using
\begin{equation}\label{eq:sumcn}
\sum_n n^{\a}\ex{A\s{n}} q^n \sim (1-z)^{-\fr{A^2}{4\pi^2}}\pa{\log (1-z)}^{2\a+\fr{3}{2}}.
\end{equation}
Inserting the expression for $A$ (\ref{eq:AABBA}), (\ref{eq:ABBAA}) into this approximation form, we can reproduce the conclusions 
 (\ref{eq:FMresult1}), (\ref{eq:FMresult2}) in Appendix \ref{subapp:FM}. In other words, our conclusions in Appendix \ref{app:FM} are supported by numerical verifications and analytic calculations in a particular case. However, we have not attempted to reproduce the logarithmic dependence in (\ref{eq:sumcn}) using the fusion matrix approach. We plan to take it up as our future work.

\clearpage
\bibliographystyle{JHEP}
\bibliography{LC}

\end{document}